\documentclass[12pt]{JHEP3}
\usepackage[centertags]{amsmath}
\usepackage{amssymb}
\usepackage{epsfig} 
\usepackage{amsmath} 
\usepackage{graphicx} 

\def\beq{\begin{equation}}
\def\eeq{\end{equation}}
\def\bea{\begin{eqnarray}}
\def\eea{\end{eqnarray}}
\def\nn{\nonumber}

\def\sss{\scriptscriptstyle}

\def\roughly#1{\mathrel{\raise.3ex\hbox
{$#1$\kern-.75em\lower1ex\hbox{$\sim$}}}}

\def\sla#1{\raise.15ex\hbox{$/$}\kern-.57em #1}

\def\ks{K_{\sss S}}

\def\bd{B_d^0}
\def\bs{B_s^0}
\def\bdbar{{\bar B}_d^0}
\def\bsbar{{\bar B}_s^0}

\def\btos{{ b} \to { s}}

\def\AFB{A_{\rm FB}}
\def\AFBq{A_{\rm FB}(q^2)}
\def\bkll{{\bar B} \to {\bar K}^* \mu^+ \, \mu^-}
\def\DBRq{dB/dq^2}
\def\btopik{B \to \pi K}

\def \kstar{{K^*}}
\def\lesssim{\mathrel{\hbox{\rlap{\hbox{\lower4pt\hbox{$\sim$}}}\hbox{$<$}}}} 
\def\gtrsim{\mathrel{\hbox{\rlap{\hbox{\lower4pt\hbox{$\sim$}}}\hbox{$>$}}}} 
\newcommand{\h}{\hat}

\title{\boldmath New-Physics Contributions to the
  Forward-Backward Asymmetry in $ B \to K^* \mu^+ \mu^-$}

\author{
Ashutosh Kumar Alok$^a$, Amol Dighe$^b$, Diptimoy Ghosh$^b$, David 
London$^a$, 
Joaquim Matias$^c$, Makiko 
Nagashima$^a$ and Alejandro Szynkman$^a$ 
\\
$^a$ Physique des Particules, Universit\'e de Montr\'eal, \\ C.P. 6128,
succ. centre-ville, Montr\'eal, QC, Canada H3C 3J7 \\
$^b$ Tata Institute of Fundamental
Research, Homi Bhabha Road, \\ Mumbai 400005, India \\
$^c$ Universitat Autonoma de Barcelona, Institut de Fisica d'Altes 
Energies, E-08193 Bellaterra, 
Barcelona, Spain \\
E-mail: \email{alok@lps.umontreal.ca},
 \email{amol@theory.tifr.res.in},
\email{diptimoyghosh@theory.tifr.res.in},
\email{london@lps.umontreal.ca},
\email{matias@ifae.es},
\email{makiko@lps.umontreal.ca},
\email{szynkman@lps.umontreal.ca}
}

\preprint{UdeM-GPP-TH-09-186,TIFR/TH/09-43,UAB-FT/675}

\abstract{
We study the forward-backward asymmetry ($\AFB$) and the differential
branching ratio (DBR) in $B \to K^* \mu^+ \mu^-$ in the presence of
new physics (NP) with different Lorentz structures.  We consider NP
contributions from vector-axial vector (VA), scalar-pseudoscalar (SP),
and tensor (T) operators, as well as their combinations.  We calculate
the effects of these new Lorentz structures in the low-$q^2$ and
high-$q^2$ regions, and explain their features through analytic
approximations. We find two mechanisms that can give a significant
deviation from the standard-model predictions, in the direction
indicated by the recent measurement of $\AFB$ by the Belle
experiment. They involve the addition of the following NP operators:
(i) VA, or (ii) a combination of SP and T (slightly better than T
alone). These two mechanisms can be distinguished through measurements
of DBR in $B \to K^* \mu^+ \mu^-$ and $\AFB$ in $B \to K
\mu^+\mu^-$.
}


\begin{document}

\section{Introduction}

To date, the standard model (SM) has been enormously successful in
explaining the measurements of particle-physics experiments.  However,
recently some discrepancies with the predictions of the SM have been
observed in $B$ decays.  Some examples are: (i) the values of the
$\bd$-$\bdbar$ mixing phase $\sin 2\beta$ obtained from different
penguin-dominated $\btos$ channels tend to be systematically smaller
than that obtained from $\bd\to J/\psi \ks$ \cite{btos}, (ii) in
$\btopik$ decays, it is difficult to account for all the experimental
measurements within the SM \cite{piKupdate}, (iii) the measurement of
the $\bs$-$\bsbar$ mixing phase by the CDF and D0 collaborations
deviates from the SM prediction \cite{BS}, (iv) the isospin asymmetry
between the neutral and the charged decay modes of the ${\bar B}\to
{\bar K}^*l^+l^-$ decay also differs from the SM \cite{fm}.  Though
the disagreements are only at the level of $\sim 2$-$3\sigma$, and
hence not statistically significant, they are intriguing since they
all appear in $\btos$ transitions.

Recently, one such discrepancy has been observed in the lepton
forward-backward asymmetry ($\AFB$) in the exclusive decay $\bkll$
\cite{Belle,BaBar}. This is especially interesting since it is a
CP-conserving process, whereas most of the other discrepancies involve
CP violation.  The deviation from the SM can be seen in the
differential $\AFB$ as a function of the dilepton invariant mass
$q^2$.  In the high-$q^2$ region ($q^2 \ge 14.4$ GeV$^2$), the
$\AFB(q^2)$ measurements tend to be larger than the SM expectations,
although both show similar trends.  The anomaly is more striking at
low $q^2$ ($1$ GeV$^2 \le q^2 \le 6$ GeV$^2$).  In the first half of
this region ($q^2 \le 3$ GeV$^2$), the SM prediction is firmly
negative \cite{Altmannshofer:2008dz}, whereas the data favor positive
values.  Moreover, the SM predicts a zero crossing in $\AFB(q^2)$
whose position is well-determined and free from hadronic uncertainties
at leading order (LO) in $\alpha_s$
\cite{zero-reference,Beneke:2001at}.  The measurements, on the other
hand, prefer positive values for $\AFB(q^2)$ in the whole $q^2$-range,
suggesting that there might not be a zero crossing.  Indeed, Belle has
claimed that this disagreement shows a clear hint of physics beyond
the SM \cite{BellePR}.

It is therefore quite natural to explore new-physics (NP) explanations
of $\AFBq$, and look for the effect of this NP on other observables in
the same decay \cite{CGS}. $\bkll$ is described by the quark-level
transition $b \to s \mu^+\mu^-$.  This is a flavor-changing
neutral-current (FCNC) process, and is therefore expected to play an
important role in the search for physics beyond the SM.  There have
already been a number of theoretical studies, both within the SM
\cite{TheobsllSM} and in specific NP scenarios \cite{TheobsllNP},
focusing on the branching fraction and $\AFB$ of $\bkll$.  For
example, Ref.~\cite{HHM} has pointed out that $\AFBq$ is a sensitive
probe of NP that affects the SM Wilson coefficients. Other observables
based on the $K^*$ spin amplitudes of this decay are at present under
active theoretical and experimental analysis
\cite{Kruger:2005ep,Lunghi:2006hc,Egede:2008uy}.  Finally, more
challenging observables, such as the polarized lepton forward-backward
asymmetry \cite{PorlAFB}, have also been considered, though the
measurement of this quantity is still lacking.

In the coming years, the LHCb experiment \cite{marco} will collect
around 6.4k events in the full range of $q^2$ for an integrated
luminosity of 2 fb$^{-1}$ (a nominal one-year data taking). This would
allow the extraction of the SM zero (if it is there) of $\AFB$ with a
precision of $\pm 0.5$ GeV$^2$. Indeed a dataset of 100 pb$^{-1}$
would already improve the world precision obtained by Babar, Belle and
CDF. These measurements would also permit many of the additional tests
for NP mentioned above.

The decays ${\bar B}\to X_s\mu^+\mu^-$ and ${\bar B}\to {\bar K}\mu^+\mu^-$ are also
described by $\btos \mu^+\mu^-$, and hence the same new physics would
be expected to affect their measurements.  The branching ratios of
these decays offer significant constraints on NP contributions from
all Lorentz structures.  The possibility of a large $\AFB$ in ${\bar B}\to
{\bar K}\mu^+\mu^-$ was considered in Ref.~\cite{ADSKmumu}, where a general
analysis, allowing for all possible NP effects, was performed.  This
included vector-axial vector (VA), scalar-pseudoscalar (SP), and
tensor (T) operators.  It was shown that $\AFBq$ in this decay cannot
be enhanced significantly with new VA operators, while T operators can
increase $\AFBq$ efficiently, especially when combined with the SP new
physics.

In this paper, we apply the method of Ref.~\cite{ADSKmumu} to the
decay $\bkll$.  That is, we perform a general analysis of NP effects
without restricting ourselves to a specific model.  Our aim here is
not to obtain precise predictions, but rather to obtain an
understanding of how the NP affects the observables, and to establish
which Lorentz structure(s) can accommodate the observed $\AFBq$
anomaly.  The impact of NP in $\AFBq$ may be partly washed out by
integrating over $q^2$, so we study the differential $\AFBq$ in the
entire $q^2$ region.

We find that, after taking into account the constraints from relevant
measurements, there are two NP Lorentz structures that can give
predictions closer to the low-$q^2$ $\AFB$ data than the SM.  The
first is the case in which one adds new VA operators.  Here, the
values of $\AFBq$ can be always positive, and hence there is no zero
crossing.  In the second, NP T operators are present, which can shift
the crossing point to much lower $q^2$ values.  The addition of SP
operators to the T operators allows the results to be somewhat
closer to the data.  We also point out the effects of viable NP
scenarios on the differential branching fraction $dB/dq^2$.

In section~\ref{bkll:SM}, we review the decay $\bkll$ within the SM. We
introduce new physics in section~\ref{bkll:NP} by adding all possible NP
operators to the effective Hamiltonian.  We also calculate the
constraints on the coefficients of these operators, and present the
theoretical expressions for $\AFBq$ and $dB/dq^2$ for
$\bkll$. Section~\ref{res_afb} contains our numerical results for $\AFBq$
and $dB/dq^2$ with the addition of specific viable NP operators.  In
section~\ref{summary}, we summarize our findings and discuss their
implications.  Some of the more complicated algebraic expressions can
be found in the appendix \ref{appendix-1}.

\section{\boldmath $\bkll$: Standard Model
\label{bkll:SM}}

Within the SM, the effective Hamiltonian for the quark-level
transition $\btos \mu^+ \mu^-$ is
\bea
{\cal H}_{\rm eff}^{SM} &~=~& -\frac{4 G_F}{\sqrt{2}}
\, V_{ts}^* V_{tb} \, \Bigl\{ \sum_{i=1}^{6} {C}_i (\mu) {\cal O}_i (\mu)
+ C_7 \,\frac{e}{16 \pi^2}\, (\bar{s} 
  \sigma_{\mu\nu} (m_s P_L + m_b P_R) b) \,
F^{\mu \nu} \nonumber \\
&& +\, C_9 \,\frac{\alpha_{em}}{4 \pi}\, (\bar{s}
\gamma^\mu P_L b) \, \bar{\mu} \gamma_\mu \mu + C_{10}
\,\frac{\alpha_{em}}{4 \pi}\, (\bar{s} \gamma^\mu P_L b) \, \bar{\mu} 
\gamma_\mu \gamma_5  
\mu  \, \Bigr\} ~,
\label{HSM}
\eea
where $P_{L,R} = (1 \mp \gamma_5)/2$. The operators ${\cal O}_i$
($i=1,..6$) correspond to the $P_i$ in Ref.~\cite{bmu}.  The SM Wilson
coefficients take the following values at the scale $\mu=4.8$ GeV in
next-to-next-to-leading order \cite{Altmannshofer:2008dz}:
\beq 
C_{7}^{\rm eff} = -0.304 ~~,~~
C_{9}^{\rm eff} = 4.211 +Y(q^2) ~~,~~ 
C_{10} = -4.103 ~,
\eeq 
where $C_7^{\rm eff}=C_7-C_3/3-4 C_4/9 -20 C_5/3 - 80 C_6/9$, $q^\mu$
is the sum of the 4-momenta of the $\mu^+$ and $\mu^-$, and the
function $Y(q^2)$ is given by \cite{Beneke:2001at}
\begin{eqnarray}
Y(q^2) &\!=\!& h(q^2,m_c) \left(\frac{4}{3} C_1 + C_2 + 6\, C_3 + 60\, 
C_5 \right) \nn\\
&&-~\frac{1}{2} h(q^2,m_b) \left(7 C_3 + \frac{4}{3} C_4 + \, 76 
C_5 + 
\frac{64}{3} C_6 \right) \\
&&-~\frac{1}{2} h(q^2,0) \left( C_3 + \frac{4}{3} C_4 + 16\, C_5 + 
\frac{64}{3} C_6 
\right) + \frac{4}{3} C_3 + \frac{64}{9} C_5 + \frac{64}{27} C_6 ~. \nn
\end{eqnarray}
Here
\begin{equation}
h(s,m_q) = -\frac{4}{9}\left(\ln\frac{m_q^2}{\mu^2} - \frac{2}{3} 
- x \right)- \frac{4}{9} \,(2+x) \,\sqrt{\,|x-1|} \,
\left\{
\begin{array}{l}
\,\arctan\displaystyle{\frac{1}{\sqrt{x-1}}}
\qquad\quad x>1\\[0.4cm]
\,\ln\displaystyle{\frac{1+\sqrt{1-x}}{\sqrt{x}}} - \frac{i\pi}{2}
\quad x\leq 1
\end{array}
\right.
\end{equation}
with $x=4 m_q^2/q^2$. A tiny weak phase has been neglected.

The decay amplitude for ${\bar B}(p_1) \to {\bar K}^*(p_2,\epsilon) \, \mu^+(p_+) \,
\mu^-(p_-)$ is 
\bea 
M(\bkll) &~=~& \frac{\alpha G_F}{2\sqrt{2} \pi} \, V_{ts}^*
V_{tb} \times \nn \\
&& \hskip-3truecm \Big[\langle {\bar K}^*(p_2,\epsilon)
\left|\bar{s}\gamma^{\mu}(1-\gamma_5)b\right|{\bar B}(p_1)\rangle
\Big\{C_{9}^{\rm eff} \, \bar{u}(p_-)\gamma_{\mu}v(p_+)
+C_{10} \, \bar{u}(p_-)\gamma_{\mu}\gamma_{5} v(p_+)\Big\}
\nn \\
& & \hskip-1.8truecm -~2 \, \frac{C^{\rm eff}_7}{q^2} \, m_b
\langle {\bar K}^*(p_2,\epsilon) | \bar{s} i\sigma_{\mu\nu} q^{\nu}
(1+\gamma_5) b |{\bar B}(p_1)\rangle \, \bar{u}(p_-) \gamma^{\mu}
v(p_+) \Big] \; ,
\eea 
where we have neglected the strange-quark mass $m_s$.  The expressions
for the matrix elements as a function of form factors are given in
Ref.~\cite{ABHH}, and are reproduced in Appendix \ref{appendix-1} for
the sake of completeness.

The double differential decay rate is given by
\beq
\frac{d^2\Gamma}{dq^2 d\cos\theta} = \frac{1}{2m_B}
\frac{2 v \sqrt{\lambda}}{(8 \pi)^3} |M|^2 \; ,
\label{ddbr}
\eeq 
where $v \equiv \sqrt{1 - 4 m_l^2/q^2}$.  Here $\lambda \equiv 1 +
\hat{r}^2 + z^2 - 2 (\hat{r} +z)-2\hat{r} z$, with $\hat{r} \equiv
m_{K^*}^2/m_B^2$ and $z \equiv q^2/m_B^2$.  The forward-backward
asymmetry for the muons is defined by
\beq
\AFBq = \frac {\displaystyle \int_{0}^{1} d\cos\theta
\frac{d^2\Gamma}{dq^2 d\cos\theta} - \int_{-1}^{0} d\cos\theta
\frac{d^2\Gamma}{dq^2 d\cos\theta} }{\displaystyle \int_{0}^{1}
d\cos\theta \frac{d^2\Gamma}{dq^2 d\cos\theta} + \int_{-1}^{0}
d\cos\theta \frac{d^2\Gamma}{dq^2 d\cos\theta} } ~,
\eeq
where $\theta$ is the angle between the momenta of the $B$ and the
$\mu^+$ in the dimuon center-of-mass frame.

\FIGURE[t]{
\epsfig{file=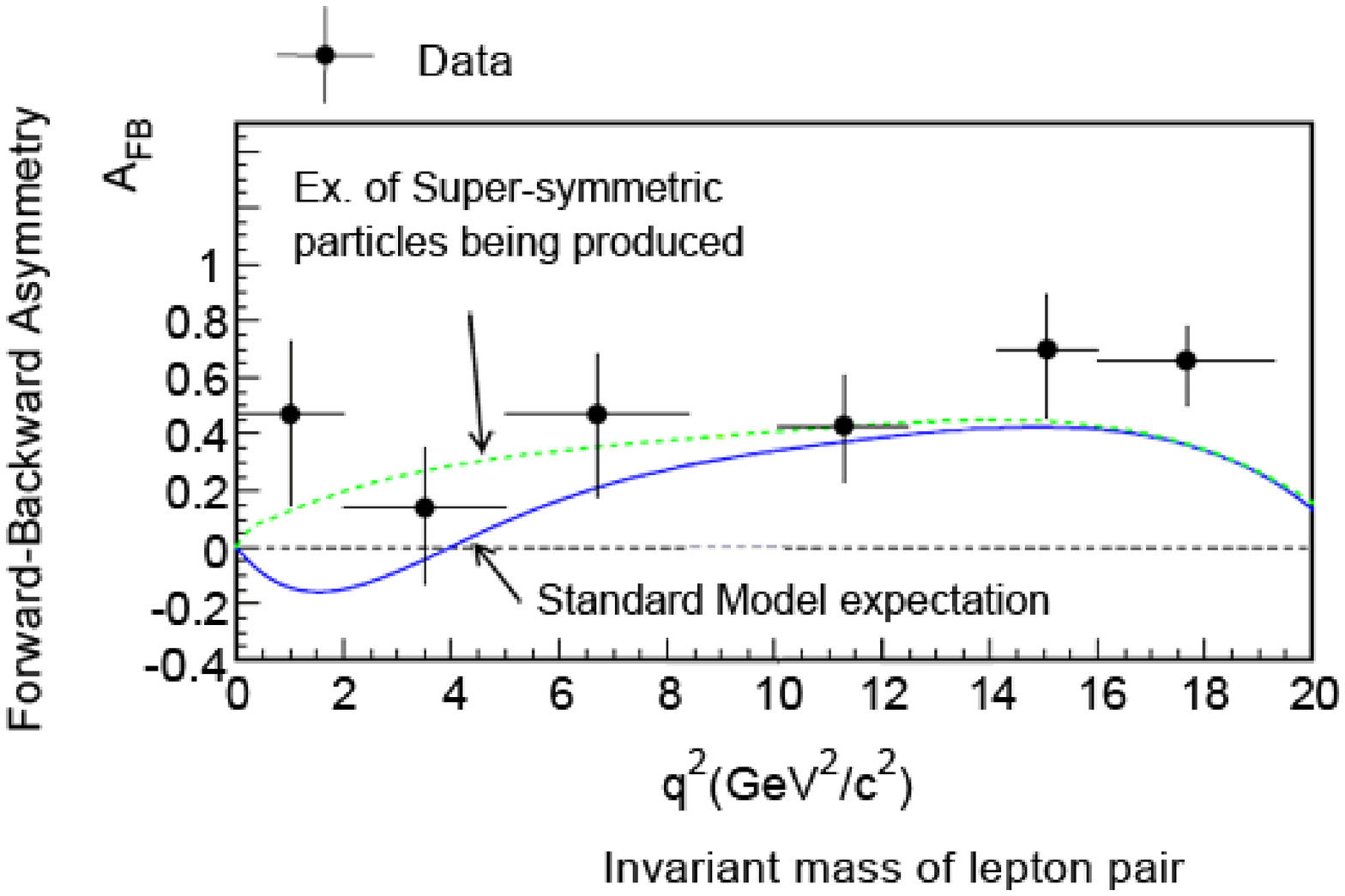,width=10cm}
\caption{The SM prediction for $\AFBq$ in $\bkll$ and the
experimental measurements from Belle.  This figure is taken
from Ref.~\cite{BellePR}.} 
\label{expt}
}

In Fig.~\ref{expt}, we show the SM prediction for $\AFBq$, along with
the experimental measurements from Belle.  From this figure, we see
that the discrepancy with the SM is the strongest in the low-$q^2$ region,
where the SM predicts negative values of $\AFBq$, as well as a zero
crossing.  The zero of $\AFBq$ is particularly clean, because at this
point the form-factor dependence cancels at LO, and a relation between
the short-distance coefficients is predicted \cite{BurasMunz}:
\begin{equation}
\label{zero} 
{\rm Re}(C_9^{\rm eff}(q_0^2))= -\frac{2 m_B m_b}{q_0^2} \, C_7^{\rm eff} ~,
\end{equation}
where $q_0^2$ is the point where $\AFB(q_0^2) = 0$.
Next-to-leading-order (NLO) contributions shift the position of this
zero to a higher value: $q_0^2=3.90\pm 0.12 \, {\rm GeV}^2$
\cite{Altmannshofer:2008dz}.  A substantial deviation from this zero
crossing point would thus be a robust signal for new physics. This can
occur if the NP affects $C_7^{\rm eff}$ and/or $C_9^{\rm eff}$, or if
it changes the above relation itself, such as by introducing new
Wilson coefficients.  The present experimental data point towards
positive values of $\AFBq$ in the entire $q^2$ region, thus favoring a
non-crossing solution.  In the following sections, we therefore look
for sources of NP which can give rise to this feature.

\section{\boldmath $\bkll$: New-Physics Lorentz Structures
\label{bkll:NP}}

\subsection{New-physics operators}

We now add new physics to the effective Hamiltonian for $\btos \mu^+
\mu^-$, so that it becomes
\beq
{\cal H}_{\rm eff}(b \to s \mu^{+} \mu^{-}) = {\cal
H}_{\rm eff}^{SM} + {\cal H}_{\rm eff}^{VA} + {\cal H}_{\rm eff}^{SP} +
{\cal H}_{\rm eff}^{T} ~,
\label{NP:effHam}
\eeq
where ${\cal H}_{\rm eff}^{SM}$ is given by Eq.~(\ref{HSM}), while
\bea
{\cal H}_{\rm eff}^{VA} &~=~& - \frac{\alpha G_F}{\sqrt{2} \pi} \,
V_{ts}^* V_{tb} \, \Bigl\{ R_V \, \bar{s} \gamma^\mu P_L b
\, \bar{\mu} \gamma_\mu \mu + R_A \, \bar{s} \gamma^\mu P_L b
\, \bar{\mu} \gamma_\mu \gamma_5 \mu \nn \\
&& \hskip2.5 truecm +~R'_V \, \bar{s} \gamma^\mu P_R b \,
\bar{\mu} \gamma_\mu \mu + R'_A \, \bar{s} \gamma^\mu P_R b
\, \bar{\mu} \gamma_\mu \gamma_5\mu \Bigr\} ~, \\
{\cal H}_{\rm eff}^{SP} &~=~& - \frac{\alpha G_F}{\sqrt{2} \pi} \,
V_{ts}^* V_{tb} \, \Bigl\{R_S ~\bar{s} P_R b ~\bar{\mu}\mu +
R_P ~\bar{s} P_R b ~ \bar{\mu}\gamma_5 \mu \nn\\
&& \hskip2.5 truecm +~R'_S ~\bar{s} P_L b ~\bar{\mu}\mu +
R'_P ~\bar{s} P_L b ~ \bar{\mu}\gamma_5 \mu \Bigr\} \;, \\
{\cal H}_{\rm eff}^{T} &~=~& - \frac{\alpha G_F}{\sqrt{2} \pi} \,
V_{ts}^* V_{tb} \, \Bigl\{C_T ~\bar{s} \sigma_{\mu \nu } b
~\bar{\mu} \sigma^{\mu\nu}\mu + i C_{TE} ~\bar{s} \sigma_{\mu
\nu } b ~\bar{\mu} \sigma_{\alpha \beta } \mu ~\epsilon^{\mu
\nu \alpha \beta} \Bigr\} \;
\eea
are the new contributions.  Here, $R_V, R_{A}, R_V', R_A', R_S, R_P,
R_S', R_P', C_{T}$ and $C_{TE}$ are the NP couplings.  For simplicity,
in our numerical analysis of the forward-backward asymmetry and the differential
branching ratio, these couplings are taken to be real. However, for
completeness, the expressions allow for a
complex-coupling analysis.

As was done in the SM case, one can turn the expression of the
effective Hamiltonian for $\btos \mu^+ \mu^-$ into a decay amplitude
for ${\bar B}(p_1) \to {\bar K}^*(p_2) \, \mu^+(p_+) \,
\mu^-(p_-)$. This amplitude is
\bea 
M(\bkll) &~=~& \frac{\alpha G_F}{2\sqrt{2} \pi} \, V_{ts}^*
V_{tb} \times \nn \\
&& \hskip-3.5truecm \bigg[\langle {\bar K}^*(p_2,\epsilon)
  |\bar{s}\gamma^{\mu}(1-\gamma_5)b|{\bar B}(p_1)\rangle \,
  \Big\{(C_{9}^{\rm eff} + R_V) \,
  \bar{u}(p_-)\gamma_{\mu}v(p_+) \nn\\
&& \hskip4truecm +~(C_{10}+R_A) \,
  \bar{u}(p_-)\gamma_{\mu}\gamma_{5} v(p_+)\Big\} \nn \\
& & \hskip-3.2truecm +~\langle {\bar K}^*(p_2,\epsilon)
|\bar{s}\gamma^{\mu}(1+\gamma_5)b|{\bar B}(p_1)\rangle \, \Big\{R'_V
\, \bar{u}(p_-)\gamma_{\mu}v(p_+) +R'_A \, \bar{u}(p_-)
\gamma_{\mu} \gamma_{5} v(p_+)\Big\} \nn \\
& & \hskip-2.5truecm -~2 \, \frac{C^{\rm eff}_7}{q^2} \, m_b
\, \langle {\bar K}^*(p_2,\epsilon) |\bar{s} i\sigma_{\mu\nu}
q^{\nu} (1+\gamma_5)b|{\bar B}(p_1)\rangle \; \bar{u}(p_-)
\gamma^{\mu} v(p_+) \nn \\
& & \hskip-2.5truecm +~\langle {\bar K}^*(p_2,\epsilon) |\bar{s}
(1+\gamma_5)b | {\bar B}(p_1) \rangle\;\left\{R_S \, \bar{u}(p_-)
v(p_+)+ R_P \, \bar{u}(p_-) \gamma_5 v(p_+) \right\} \nn\\
& & \hskip-2.5truecm +~\langle {\bar K}^*(p_2,\epsilon) |\bar{s}
(1-\gamma_5)b | {\bar B}(p_1) \rangle\;\left\{R'_S \, \bar{u}(p_-)
v(p_+)+ R'_P \, \bar{u}(p_-)\gamma_5 v(p_+)\right\} \nn\\
& & \hskip-2.5truecm +~2 \, C_T \, \langle {\bar K}^*(p_2,\epsilon)
|\bar{s} \sigma_{\mu\nu}b | {\bar B}(p_1) \rangle \; \bar{u}(p_-)
\sigma^{\mu\nu} v(p_+) \nn \\
& & \hskip-2.5truecm +~2i \, C_{TE} \, \epsilon^{\mu \nu
  \alpha\beta} \langle {\bar K}^*(p_2,\epsilon) | \bar{s}
  \sigma_{\mu\nu} b|{\bar B}(p_1) \rangle\; \bar{u}(p_-)
  \sigma_{\alpha \beta}v(p_+) \bigg] ~,
\label{bkllampNP}
\eea 
where the expressions for the matrix elements \cite{ABHH} are
reproduced in appendix \ref{appendix-1}.  Note that the matrix
elements are functions of 7 form factors: $A_{0,1,2}(q^2)$, $V(q^2)$,
$T_{1,2,3}(q^2)$.

\subsection{Constraints on the new-physics couplings}
\label{constraint}

The constraints on the NP couplings in $b \to s \, \mu^+ \, \mu^-$ are
obtained mainly from the related decays ${\bar B} \to X_s \, \mu^+ \,
\mu^-$ and ${\bar \bs} \to \mu^+ \, \mu^-$. Due to the large hadronic
uncertainties, the exclusive decays ${\bar B} \to ({\bar K},{\bar
  K}^*)\, \mu^+ \, \mu^-$ provide weaker constraints than the
inclusive decay ${\bar B} \to X_s \, \mu^+ \, \mu^-$.

Neglecting the muon and strange-quark masses, the branching
ratio of ${\bar B} \to X_s \, \mu^+ \, \mu^-$ is given by
\bea
B({\bar B} \to X_s \, \mu^+ \, \mu^-) &~=~& B_{SM} + B_{
VA}\left[ |R_V|^2 + |R_A|^2 + {|R'_V|}^{\hskip-0.2truemm 2} + {|R'_A|}^{\hskip-0.2truemm 2}
\right] \nn\\
&& \hskip-1truecm +~B_{SM{\hbox{-}}VA} {\rm Re}\left[R_V^* C_9^{\rm eff} 
+ R_A ^* C_{10} \right] + B'_{SM{\hbox{-}}VA} {\rm Re}(R_V^* C^{\rm
eff}_7) \\
&& \hskip-1truecm +~B_{SP}\left[ |R_S|^2 + |R_P|^2 +
{|R'_S|}^{\hskip-0.2truemm 2} + {|R'_P|}^{\hskip-0.2truemm 2} \right] + 
B_{T}\left[ |C_T|^2 +
4|C_{TE}|^2 \right]\;, \nn
\label{BBXll}
\eea
where
\bea
B_{SM} &~=~& B_0 \int_{z_{\rm min}}^{z_{\rm max}} dz\;
(1-z) \Bigg[ \frac{16 }{z}\left\{1-z^2 + \frac{(1-z)^2}{3}
\right\} (C^{\rm eff}_7)^2 \nn\\
&& +~4\left\{1-z^2- \frac{(1-z)^2}{3} \right\} \left[
|C_9^{\rm eff}|^2 + C_{10}^2 \right] +32\,(1-z)\, C^{\rm
eff}_7\, {\rm Re}(C_9^{\rm eff}) \Bigg]\;, \nn \\
B_{VA} &~=~& 4\,B_0 \int_{z_{\rm min}}^{z_{\rm max}} dz\;
(1-z) \left\{1-z^2- \frac{(1-z)^2}{3} \right\} \;, \nn \\ 
B_{SM{\hbox{-}}VA} &~=~& 2\, B_{VA}\;, \qquad B_{
T} \;= \; 16\, B_{VA}\;, \nn \\
B'_{SM{\hbox{-}}VA} &~=~& 32\,B_0 \int_{z_{\rm
min}}^{z_{\rm max}} dz\; (1-z)^2\;, \qquad B_{SP} \;= \;
4\,B_0 \int_{z_{\rm min}}^{z_{\rm max}} dz\; z\,(1-z)^2\; ,
\eea 
with $z \equiv q^2/m_{b}^2$. 
The normalization constant $B_0$ is
\begin{equation}
B_0= \frac{3\alpha^2 \, B({\bar B}\rightarrow X_c e {\bar \nu})}
 {32 \pi^2 \, f(\hat{m_c}) \, \kappa(\hat{m}_c)} 
 \frac{|V_{tb}^{*}V_{ts}|^2}{|V_{cb}^{*}|^2}\;,
\end{equation}
where $\hat{m}_c \equiv m_c/m_b$. Here $f(\hat{m_c})$ is the
phase-space factor in $B({\bar B} \to X_c e {\bar \nu})$
\cite{Nir:1989rm}:
\beq
f(\hat{m}_c) = 1 - 8\hat{m}^2_c + 8\hat{m}_c^6 - \hat{m}_c^8
- 24\hat{m}_c^4 \ln \hat{m}_c \; ,
\eeq
and $\kappa(\hat{m_c})$ is the $1$-loop QCD correction factor
\cite{Nir:1989rm}
\beq
\kappa(\hat{m_c})=1-\frac{2\alpha_s(m_b)}{3\pi} \left[ \left(
  \pi^2-\frac{31}{4}\right)(1-\hat{m_c})^2 +
  \frac{3}{2}\right]\;.
\eeq

The branching ratio of ${\bar B} \to X_s \, \mu^+ \, \mu^-$ has been
measured by both Belle~\cite{Iwasaki:2005sy} and
BaBar~\cite{Aubert:2004it}.  In the low-$q^2$ ($1$ GeV$^2 \le q^2 \le
6$ GeV$^2$) and high-$q^2$ ($14.4$ GeV$^2 \le q^2 \le 25$ GeV$^2$)
regions, the measurements are
\bea
{B} ( {\bar B} \to X_s \, \mu^+ \, \mu^-)_{{\rm low}~q^2} &~=~& 
\left\{ \begin{array}{ll}
\left( 1.49 \pm 0.50^{+0.41}_{-0.32} \right) \times
10^{-6}~, & (\rm Belle)~, \\
\left( 1.8 \pm 0.7 \pm 0.5 \right) \times 10^{-6} ~, & (\rm
BaBar)~, \\
\left( 1.60 \pm 0.50 \right) \times 10^{-6}~, & (\rm Average)
~. \\
\end{array} \right. \\
{B} ( {\bar B} \to X_s \, \mu^+ \, \mu^-)_{{\rm high}~q^2} &~=~& 
\left\{ \begin{array}{ll}
\left( 0.42 \pm 0.12^{+0.06}_{-0.07} \right) \times
10^{-6} ~, & (\rm Belle) ~, \\
\left( 0.50 \pm 0.25 ^{+0.08}_{-0.07} \right) \times 10^{-6}
~, & (\rm BaBar) ~, \\
\left( 0.44 \pm 0.12 \right) \times 10^{-6} ~, 
& (\rm Average) ~. \\ 
\end{array} \right.
\eea
The SM predictions for $B({\bar B} \to X_s \, \mu^+ \, \mu^-)$ in the
low- and high-$q^2$ regions are $(1.59\pm0.11) \times 10^{-6}$ and
$(0.24 \pm 0.07) \times 10^{-6}$, respectively \cite{Huber:2007vv}.

The branching ratio of ${\bar \bs} \to \mu^+ \, \mu^-$ in the presence
of the NP operators is
\begin{eqnarray}
  B({\bar B_s} \to \mu^+ \, \mu^-) & =& \frac{G^2_F \alpha^2 m^5_{B_s} f_{B_s}^2 \tau_{B_s}}{64 \pi^3}
     |V_{tb}^{}V_{ts}^{\ast}|^2 \sqrt{1 - \frac{4 m_\mu^2}{m_{B_s}^2}} 
\nonumber\\
  && \hskip-2.5truecm \times \Bigg\{ 
    \Bigg(1 - \frac{4m_\mu^2}{m_{B_s}^2} \Bigg) \Bigg|
\frac{R_S - R'_S}{m_b + m_s}\Bigg|^2 
    + \Bigg|\frac{R_P - R'_P}{m_b + m_s} + \frac{2 m_\mu}{m^2_{B_s}} (C_{10}+R_A-R'_A)\Bigg|^2 \Bigg\}. \phantom{space}
\end{eqnarray}
The SM prediction for $B({\bar \bs} \to \mu^+ \, \mu^-)$ is $(3.35\pm
0.32)\times 10^{-9}$ \cite{blanke}.  The CDF experiment has reported
an upper bound on this branching ratio of $4.47 \times 10^{-8}$ at
90\% C.L.  \cite{cdf07}.

These two decays provide complementary information about the NP
operators.  The contribution of the SP couplings to ${\bar B} \to X_s
\mu^+ \mu^-$ is suppressed by the small coefficient $B_{SP}\sim
10^{-9}$, as compared to $B_{SM} \sim 10^{-6}$.  As a result, the
constraints on the SP coefficients from this decay are rather weak.
On the other hand, the coefficient of the tensor couplings, $B_{T}$,
is an order of magnitude larger than $B_{SP}$, while the VA operators
interfere with those of the SM ($B_{SM{\hbox{-}}VA}$).  Therefore,
this decay is sensitive mainly to the new VA and T couplings.  In
contrast, the main contributions to ${\bar \bs} \to \mu^+\mu^-$ are
precisely from the SP operators: there is no contribution from the
vector couplings $R^{(')}_{V}$, the axial-vector contribution
proportional to $R^{(')}_{A}$ is suppressed by $m_\mu/m_{B_s}$, and
there is no tensor piece since $\langle 0|\bar{s} \sigma_{\mu \nu}
b|B_s^0 (p) \rangle$ vanishes.

The constraints on the new VA couplings coming from $B({\bar B} \to
X_s \, \mu^+ \, \mu^-)$ involve the interference terms between the SM
and the NP. When $R_V$ and $R_A$ are constrained to be real, the allowed region in the $R_V$-$R_A$ parameter space
therefore looks like an annulus, as shown in the left panel of
Fig.~\ref{fig:vabounds}, as long as no other NP couplings are present.
When the couplings $R_V'$ and $R_A'$ are also permitted to be nonzero real numbers,
the allowed region takes the form of an elliptical disc, as shown in
the right panel of Fig.~\ref{fig:vabounds}.  The $R'_{V,A}$ couplings
do not interfere with the SM, so their constraints take the form of an
elliptical disc in the $R'_V$-$R'_A$ plane. If $R_{V,A}$ are not
present, the constraints are approximately
\beq
|R'_V|^2 + |R'_A|^2 \le 16.8 \; ,
\eeq
while if $R_{V,A}$ are allowed, these constraints are somewhat
weakened to
\beq
|R'_V|^2 + |R'_A|^2 \le 39.7 \; .
\eeq


\FIGURE[t]{
\epsfig{file=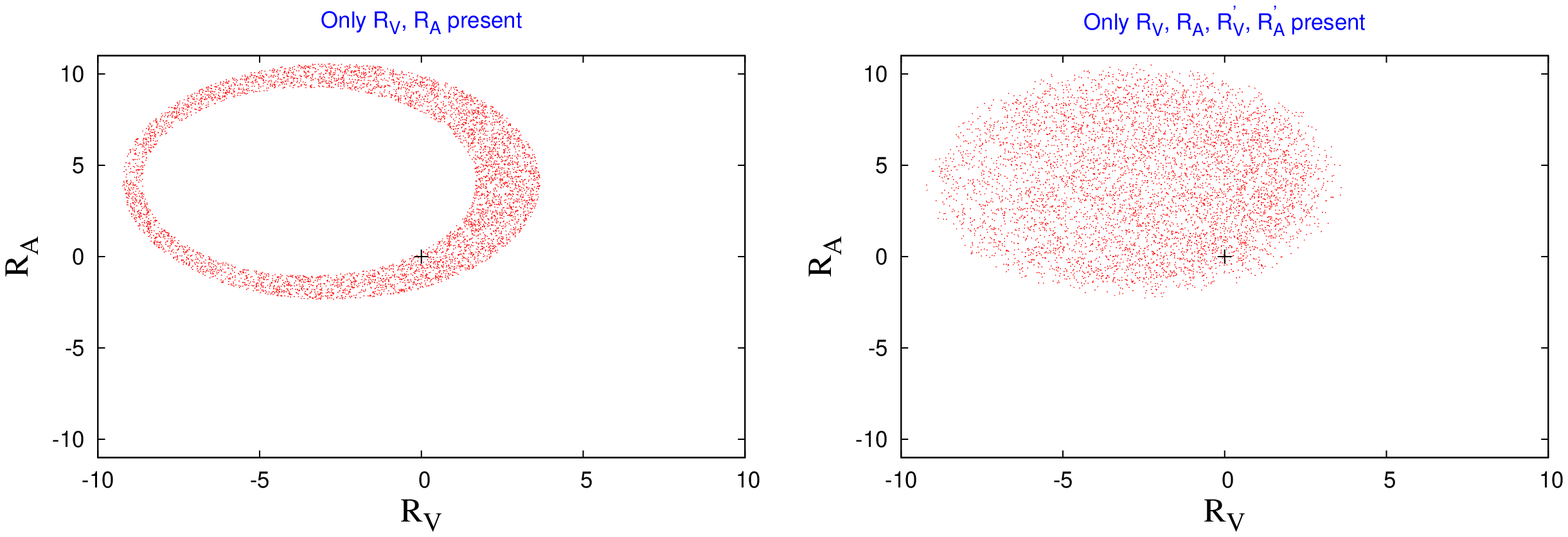,width=15cm}
\caption{Allowed parameter space in the $R_V$-$R_A$ plane when
  $R'_{V,A}$ couplings are absent (left panel) and present
  (right panel). All the couplings have been taken to be real.
The ``+'' corresponds to the SM.}
\label{fig:vabounds}
}

The constraints on the tensor operators also come entirely from
$B({\bar B} \to X_s \, \mu^+ \, \mu^-)$ and are rather tight. We find
that the allowed values of the new tensor couplings are restricted to
\begin{equation}
|C_T|^2 +4 |C_{TE}|^2 \le 1.3 ~.
\end{equation}

For the SP operators, the present upper bound on $B({\bar \bs} \to \mu^+ \,
\mu^-)$ gives the limit
\begin{equation}
|R_S - R'_S|^2 + |R_P - R'_P|^2 \le 0.44 \; ,
\end{equation}
where we have used $|V_{ts}|=(0.0407 \pm 0.0010)$ \cite{pdg} and
$f_{B_s}=(0.243\pm0.011)$ GeV \cite{Aubin:2009yh}. If only $R_{S,P}$
or $R'_{S,P}$ are present, this constitutes a severe constraint on the
NP couplings. However, if both types of operators are present, these
bounds can be evaded due to cancellations between the $R_{S,P}$ and
$R'_{S,P}$.  In that case, the constraints on these couplings come
mainly from $B({\bar B} \to X_s \, \mu^+ \, \mu^-)$, and are rather weak:
\beq
|R_S|^2 + |R_P|^2 < 45 \; , \quad R'_S = R_S \; , \quad R'_P = R_P \; .
\eeq
However, we shall ignore such fine-tuned situations.

\subsection{Forward-backward asymmetry and the 
differential branching ratio}
\label{afb-dbdqsq}

The double differential decay rate $d^2\Gamma/dq^2 d\cos\theta$,
calculated by substituting the matrix element from
Eq.~(\ref{bkllampNP}) into Eq.~(\ref{ddbr}), in turn leads to the
calculation of $dB/dq^2$ and $\AFBq$.

The differential branching ratio is given by
\begin{equation}
\frac{dB}{dq^2} = \frac{G^2 \alpha^2}{2^{14}} \frac{1}{\pi^5}
|V_{tb}V_{ts}^{*}|^2 m_{B} \tau_B \sqrt{\lambda} \, \Theta ~,
\end{equation}
where $\tau_B$ is the lifetime of $B$ meson.  The quantity $\Theta$
has the form
\beq
\Theta = \frac{1}{3\h{r}}\Big[X_{SP}+X_{VA}+X_{T}+X_{SP{\hbox{-}}VA}+X_{SP{\hbox{-}}T}
+X_{VA{\hbox{-}}T}\Big] \; ,
\label{theta-expansion}
\eeq
where the $X$ terms are classified according to the contributions they
contain coming from scalars-pseudoscalars, vectors-axial vector and
tensor operators.  Their complete expressions are given in Appendix
\ref{appendix-1}.  Note that the SM contribution is contained inside
the $X$ terms labeled by VA.  Therefore whenever the new VA
operators are absent, the $X_{VA},X_{SP{\hbox{-}}VA},
X_{VA{\hbox{-}}T}$ terms will be referred to as
$X_{SM},X_{SP{\hbox{-}}SM}, X_{SM{\hbox{-}}T}$, respectively, for
clarity.

The forward-backward asymmetry can also be written in the form
\beq 
\AFBq =2 m_{B} \frac{\sqrt{\lambda}}{\hat{r} \Theta}
\Big[Y_{SP}+Y_{VA}+Y_{T}+Y_{SP{\hbox{-}}VA}+Y_{SP{\hbox{-}}T}+Y_{VA{\hbox{-}}T}\Big]
\;,
\label{afb-expansion}
\eeq
with the complete expressions for the $Y$ terms given in Appendix
\ref{appendix-1}.  As in the case of the $X$ terms, whenever new VA
operators are absent, we refer to the $Y_{VA},Y_{SP{\hbox{-}}VA},
Y_{VA{\hbox{-}}T}$ terms as $Y_{SM},Y_{SP{\hbox{-}}SM},
Y_{SM{\hbox{-}}T}$, respectively.

Most of the qualitative features of the NP impact on the above
quantities can be easily understood if we use simplified expressions
that neglect terms proportional to the small quantities $\hat{m}_l$
and $\hat{r}$ at appropriate places. (Note that this may not be valid
for extremely low values of $q^2$.)  With this approximation, the
terms in $dB/dq^2$ simplify to
\begin{eqnarray}
\label{X-simplified}
X_{SP} & \approx & 3 (|B_1|^2 + |B_{2}|^2) m_{B}^2 z  \lambda  \; , \nn \\
X_{VA} & \approx & 
2 (|C|^2 + |G|^2) m_B^2 \lambda^2 
+ 2 ( |B|^2 + |F|^2) (12 \h{r} z + \lambda) \nn\\
&& \hskip4truecm
-~4 {\rm Re}(F G^*+ BC^*) m_B^2 (1-z) \lambda \; , \phantom{spa} \nn\\
X_{T} & \approx& |C_{T}|^2 ({\rm Quadratic~ terms~ in~} B_3, B_4, T_1) \nn\\
&& \hskip4truecm
+ |C_{TE}|^2 ({\rm Quadratic~ terms~ in~} B_3, B_4, T_1) \; .
\end{eqnarray}
The three interference terms, $X_{SP{\hbox{-}}VA}, X_{SP{\hbox{-}}T}$
and $X_{VA{\hbox{-}}T}$ vanish in this approximation, indicating that
$dB/dq^2$ can be thought of as the simple addition of the SP, VA, and
T contributions.

With the same approximations, the only surviving $Y$ terms in $\AFBq$
are
\begin{eqnarray}
Y_{VA}&\approx& -4 m_{B} \h{r} z {\rm Re}\Big(A^* F + B^* F_1\Big) \nonumber \\ 
Y_{SP{\hbox{-}}T}&\approx& m_B z {\rm Re}\Big(2 B_{1}^* C_{TE} +  B_{2}^* C_T \Big) \times \nn\\
&& \hskip3truecm
\Big((2 B_{3}-4T_1) (z-1)+B_{4} m_{B}^2 \lambda \Big)  \; .
\label{Y-simplified}
\end{eqnarray}
The chiral structure of the operators ensures that all the other 
terms are suppressed by $\hat{m}_l=m_l/m_B \approx 0.02$.

The approximate expressions in Eqs.~(\ref{X-simplified}) and 
(\ref{Y-simplified}) imply that
\begin{itemize}

\item New interactions of the type only SP or only T only always increase
  $\Theta$, and hence $dB/dq^2$, but do not contribute to $Y$.  As a
  result, $\AFBq$ always decreases in magnitude with such new physics.

\item New VA interactions, or an SP-T combination, is required in
  order to enhance $\AFBq$ significantly, or to change its sign.

\end{itemize}

Note that the SP-T contribution was already considered in
Ref.~\cite{FKMY} in the context of the inclusive decay ${\bar B} \to
X_s l^+ l^-$.  However, it was concluded that its effect was basically
to increase the branching ratio while leaving unchanged the integrated
$\AFB$.  For this reason, this contribution was disregarded in
subsequent papers such as Ref.~\cite{AKK}.  However, as we shall show
here, this type of NP can in fact shift the differential asymmetry
$\AFBq$ towards the Belle data.

In order to determine the numerical values of $dB/dq^2$ and $\AFBq$,
we need to calculate the form factors.  The theoretical predictions
for $\AFBq$ are rather uncertain in the intermediate region
($7$~GeV$^2 \le q^2 \le 12$~GeV$^2$) due to nearby charmed resonances.
The predictions are relatively more robust for lower and higher $q^2$.
We therefore concentrate on calculating $\AFBq$ in the low-$q^2$
($1~{\rm GeV^2} \le q^2 \le 6~{\rm GeV^2}$) and the high-$q^2$ ($q^2
\ge 14.4~{\rm GeV^2}$) regions.

\subsubsection{Form factors in the low-$q^2$ region}

When the initial hadron contains the heavy $b$ quark, and the final
meson has a large energy, the hadronic form factors can be expanded in
the small ratios $\Lambda_{\rm QCD}/m_b$ and $\Lambda_{\rm QCD}/E$,
where $\Lambda_{\rm QCD}$ is the strong interaction scale and $E$ is
the energy of the light meson. Neglecting corrections of
$O(\alpha_s)$, the 7 a-priori independent $B \to K^*$ form factors
[see Eqs.~(\ref{me1})--(\ref{me4})] can be expressed in terms of two
universal form factors $\xi_{\bot}(q^2)$ and $\xi_{\|}(q^2)$
\cite{Charles:1998dr, Charles:1999gy, Dugan:1990de, Beneke:2000wa}:
\bea
      A_1(q^2) &~=~& \frac{2 E_\kstar}{m_B + m_\kstar} \, \xi_{\bot}(q^2)\;, \nonumber \\
      A_2(q^2) &~=~& \frac{m_B}{m_B-m_\kstar} \bigg[\xi_{\bot}(q^2)- \xi_{\|}(q^2)\bigg]\;, \nonumber \\
      A_0(q^2) &~=~&\frac{E_\kstar}{m_\kstar} \, \xi_{\|}(q^2)\;, \nonumber \\
      V(q^2) &~=~& \frac{m_B+m_\kstar}{m_B} \, \xi_{\bot}(q^2)\;, \nonumber \\
      T_1(q^2) &~=~& \xi_{\bot}(q^2)\;, \nonumber \\
      T_2(q^2) &~=~& \frac{2 E_\kstar}{m_B} \, \xi_{\bot}(q^2)\;, \nonumber \\
      T_3(q^2) &~=~& \xi_{\bot}(q^2) - \xi_{\|}(q^2)\;.
\eea
\label{form:factor:relations:LEL}
Here, $E_\kstar$ is the energy of the $\kstar$ in the ${B}$
rest frame:
\beq
E_\kstar \simeq  \frac{m_B}{2}\left(1-\frac{q^2}{m_B^2} \right).
\eeq
The $q^2$-dependence of the form factors is assumed to be
\cite{Beneke:2001at}
\beq
\xi_{\|}(q^2) = \xi_{\|}(0) \left[ \frac{1}{1-q^2/m^2_B}
\right]^2 ~~,~~~~
\xi_{\bot}(q^2) = \xi_{\bot}(0) \left[ \frac{1}{1-q^2/m^2_B}
\right]^3 ~,
\eeq
as predicted by power counting in the heavy-quark limit. In
our analysis, we take \cite{Beneke:2001at}
\beq
\xi_{\parallel}(0) = 0.16 \pm 0.03 ~~,~~~~ \xi_{\perp}(0)=
0.26 \pm 0.02 ~.
\eeq
The previous relations get corrections of $O(\alpha_s)$
\cite{Beneke:2001at} and possible $\Lambda/m_b$
contributions. However, for our analysis it is sufficient to stay at
LO to determine which new couplings can induce a clear change of
behavior of $\AFBq$.

\subsubsection{Form factors in the high-$q^2$ region}

In order to estimate $A_{FB}(q^2)$ in the high-$q^2$ region,
we take the form factors calculated in the QCD sum rule
approach \cite{ABHH}.  The $z~(\equiv q^2/m^2_B)$ dependence
of the 7 form factors is given by
\begin{eqnarray} 
f(z)=f(0)\,\exp(c_1z+c_2z^2) ~.
\label{qdep}
\end{eqnarray} 
The central values of the parameters $f(0)$, $c_1$ and $c_2$ for each form factor
are given in Table~\ref{ff-table1}. In order to take into account form factor uncertainties, 
we have used the maximum and minimum allowed values of the parameters  $f(0)$, $c_1$ and $c_2$ as given in 
\cite{ABHH}.\\
\TABLE[t]{
\label{ff-table1}
\begin{tabular}{lcccccc}
\hline
\hspace{0.7cm} & $\phantom{-}f(0)$ \hspace{0.7cm} &$\phantom{-}c_1  $\hspace{0.7cm} & $\phantom{-}c_2$ \\ 
\hline 
 $A_1$ \hspace{0.7cm} & $\phantom{-}0.337 $ \hspace{0.7cm} & $\phantom{-}0.602$ \hspace{0.7cm} & $\phantom{-}0.258$ \\ 
$A_2 $ \hspace{0.7cm} & $\phantom{-}0.282 $ \hspace{0.7cm} & $\phantom{-}1.172  $ \hspace{0.7cm} & $\phantom{-}0.567 $ \\ 
$A_0 $ \hspace{0.7cm} & $\phantom{-}0.471 $ \hspace{0.7cm} & $\phantom{-}1.505 $ \hspace{0.7cm} & $ \phantom{-}0.710 $\\ 
$V $ \hspace{0.7cm} & $\phantom{-}0.457 $ \hspace{0.7cm} & $\phantom{-}1.482 $ \hspace{0.7cm} & $ \phantom{-}1.015  $\\ 
$T_1 $ \hspace{0.7cm} & $\phantom{-}0.379 $ \hspace{0.7cm} & $\phantom{-}1.519 $ \hspace{0.7cm} & $ \phantom{-}1.030 $\\ 
$T_2 $ \hspace{0.7cm} & $\phantom{-}0.379 $ \hspace{0.7cm} & $\phantom{-}0.517 $ \hspace{0.7cm} & $ \phantom{-}0.426 $\\ 
$T_3 $ \hspace{0.7cm} & $\phantom{-}0.260 $ \hspace{0.7cm} & $\phantom{-}1.129 $ \hspace{0.7cm} & $ \phantom{-}1.128 $\\ 
\hline 
\end{tabular}
\caption{ Central values of the parameters of the form factors for the $B \to K^*$ transition [see Eq.~(\protect \ref{qdep})] \cite{ABHH}  } 
}

\section{\boldmath $\AFBq $ and $ dB/dq^2 $ in the Presence of NP
\label{res_afb}}
In this section, we examine the predictions for $\AFBq$ and $\DBRq$ in
the presence of NP operators.  We consider different Lorentz
structures of NP, as well as their combinations, and examine the
implications using the constraints on the new couplings obtained in
Sec.~\ref{constraint}. In all figures, we show $\AFBq$ and $\DBRq$ for
representative values of the NP couplings. The representative values
have been chosen such that the maximum and minimum allowed values of
$\AFBq$ and $\DBRq$, as well as cases with interesting variations of
$\AFBq$, are displayed.  The same color (type) of line in all four
panels of a figure corresponds to the same values of NP parameters. In
addition, for comparison, we also show the experimental data. For this numerical analysis, 
we have taken the NP couplings to be real.
\subsection{VA new-physics operators}

{}From the discussion following Eq.~(\ref{Y-simplified}), it is
expected that NP in the form of vector-axial vector operators may be
able to enhance $\AFBq$ or change its sign. However, depending on
whether the NP couplings are $R_{V,A}$ or $R'_{V,A}$, the effect on
$\AFBq$ will have different features. In this section, we shall
sequentially consider the scenarios in which (i) only $R_{V,A}$
couplings are present, (ii) only $R'_{V,A}$ couplings are present, and
(iii) both types of couplings are allowed.

\subsubsection{Only $R_V$, $R_A$ couplings present}
\label{RvRa}

\FIGURE[t]{
\epsfig{file=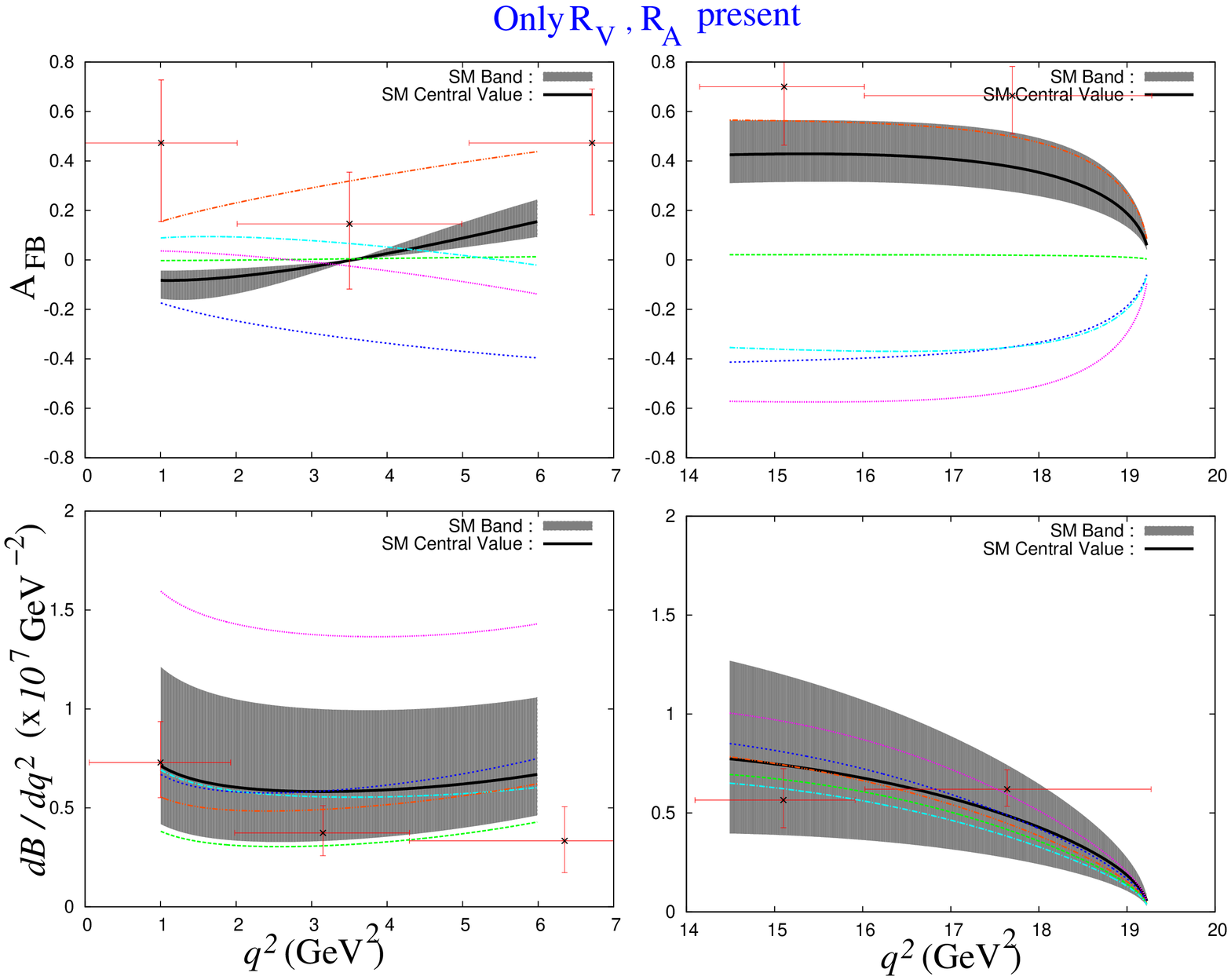,width=15cm}
\caption{The left (right) panels of the figure show $\AFBq$ and
  $\DBRq$ in the low-$q^2$ (high-$q^2$) region, in the scenario where
  only $R_V$ and $R_A$ couplings are present. The different (colored)
  curves correspond to different choices of the $R_V$ and $R_A$
  couplings inside their allowed region. For comparison, the
  experimental data are also displayed.}
\label{plot:va}
}

Fig.~\ref{plot:va} shows the results when the only NP couplings
present are $R_V$ and $R_A$. The following remarks are in order:
\begin{itemize}

\item For certain values of $R_V$ and $R_A$, $\AFBq$ can be either
  always positive (a possible solution for the Belle observation) or
  always negative.  That is, for these cases there is no zero crossing
  point.  This is easily explained because, in the presence of $R_V$
  and $R_A$, Eq.~(\ref{zero}) becomes at LO
\begin{equation} 
{\rm Re}(C_9^{\rm eff}(q_0^2)) + R_V = -\frac{2 m_B m_b}{q_0^2}
C_7^{\rm eff} ~.
\end{equation}
Then $R_V$ can unbalance the contribution from
$C_9^{\rm eff}$, so that there is no solution, and consequently
no zero.  The effect of $R_A$ is simply to rescale $\AFB$.

\item In general the zero crossing can be anywhere in the
whole $q^2$ range. The crossing can be negetive to positive
(positive crossing) or positive to negative(negative
crossing).

\item It is possible to have a large $\AFBq$ while being consistent
  with the SM prediction of the differential branching ratio
  $\DBRq$. This is explained by the different type of contributions
  entering the $X$ and $Y$ terms in Eqs.~(\ref{X-simplified}) and
  (\ref{Y-simplified}).

\item The differential branching ratio $\DBRq$ can be increased in the
  low-$q^2$ region by up to 50\%.  However, in such cases, $\AFBq$
  becomes highly negative at high $q^2$, inconsistent with the current
  data.  This suggests that, in general, $\DBRq$ will not be affected
  in this scenario.

\end{itemize}

\subsubsection{Only $R'_V$, $R'_A$ couplings present}
\label{Rv'Ra'}

\FIGURE[t]{
\epsfig{file=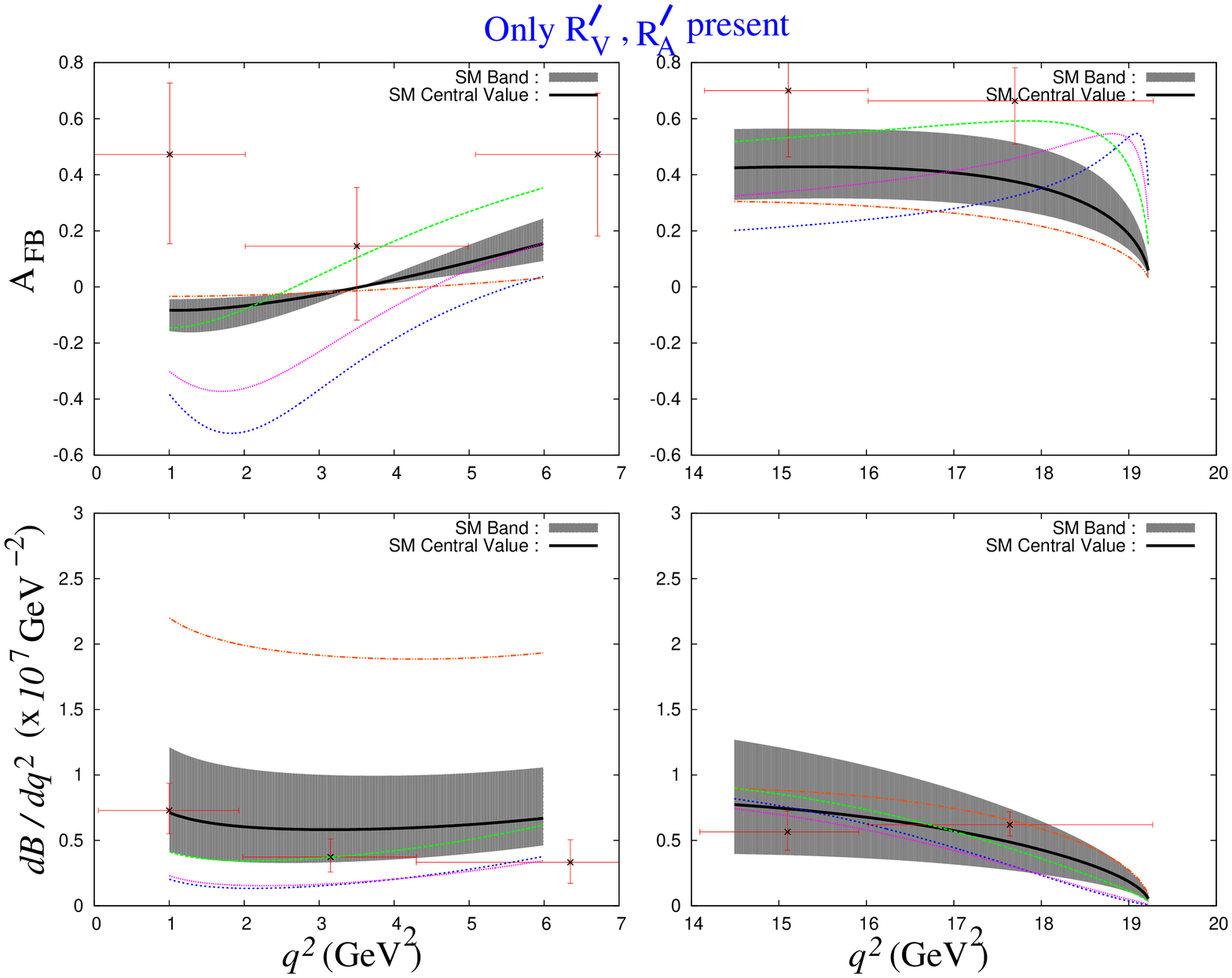,width=15cm}
\caption{The left (right) panels of the figure show $\AFBq$ and
     $\DBRq$ in the low-$q^2$ (high-$q^2$) region, in the scenario
     where only $R'_V$ and $R'_A$ terms are present.}
    \label{plot:vap}
}

Fig.~\ref{plot:vap} shows the results when the only NP couplings
present are $R'_V$ and $R'_A$. From the figure, we make the following
observations:
\begin{itemize}

\item For certain values of $R'_V$ and $R'_A$, the position of the
  zero crossing is shifted significantly, but it is always a positive
  crossing, since $\AFBq$ is highly negative in the low-$q^2$ region.
  This behavior can be understood from Eq.~(\ref{zero}), which in the
  presence of $R'_V$ and $R'_A$ becomes at LO
\begin{equation} 
Re(C_9^{\rm eff}(q_0^2)) -\frac{R'_V R'_A}{C_{10}}= - \frac{2 m_B 
m_b}{q_0^2} C_7^{\rm eff} ~.
\end{equation}
In order to counteract the contribution from $C_9^{\rm eff}$, we must
have $\left| R'_V R'_A/C_{10}\right|> {\rm Re}(C_9^{\rm eff})$.
However, this is not allowed by the present measurement of the
branching ratio of ${\bar B} \to X_s\mu^+\mu^-$.  Hence, the zero
crossing is always SM-like, i.e. always positive, which is not favored
by the Belle data.

\item It is possible to have $\DBRq$ consistent with the SM,
  simultaneously with a larger $\AFBq$ than the SM (up to 0.6), but
  only near the high-$q^2$ end.

\item $\DBRq$ at low $q^2$ can be enhanced by up to a factor of 2, but
  then $\AFBq$ would become very small. On the other hand, $\DBRq$ at
  low $q^2$ can decrease by up to 50\%, but this would result in a
  large negative value of $\AFBq$ in this region.

\end{itemize}

\subsubsection{All VA couplings present}
\label{vavap}

\FIGURE[t]{
\epsfig{file=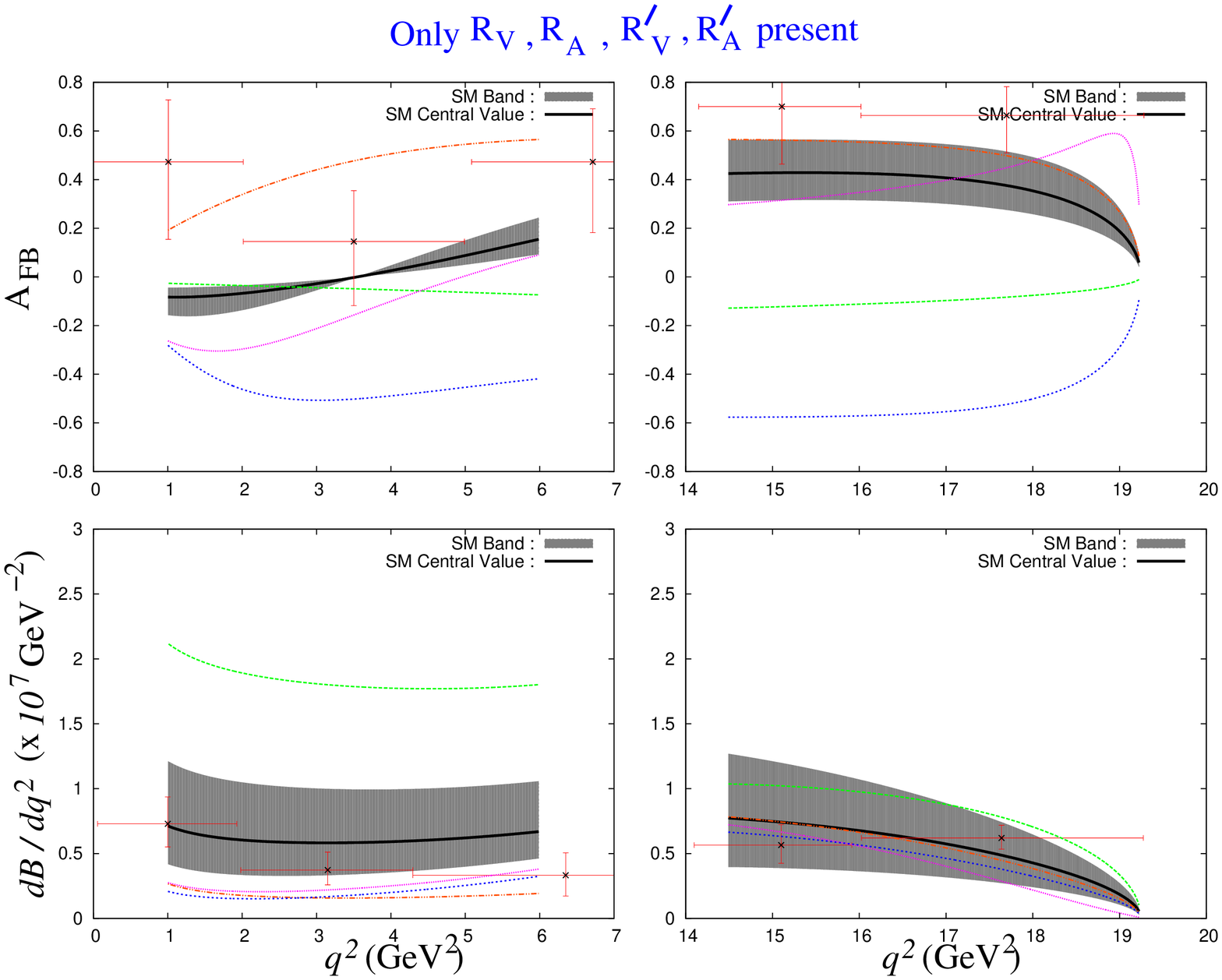,width=15cm}
\caption{The left (right) panels of the figure show $\AFBq$ and
  $\DBRq$ in the low-$q^2$ (high-$q^2$) region, in the scenario where
  both $R_{V,A}$ and $R'_{V,A}$ terms are present.}
\label{plot:vavap}
}

Fig.~\ref{plot:vavap} shows $\AFBq$ and $\DBRq$ when all the VA NP
couplings, $R_V, R_A, R'_V, R'_A$ are present. The following different
results are obtained depending on the choice of the couplings:
\begin{itemize}

\item For certain values of the couplings, $\AFBq$ can
be either always positive or always negative.  That is, there
is no zero crossing point.

\item The zero crossing can be anywhere in the whole $q^2$
range. It can be either positive or negative.

\item Particularly interesting is the case of the top curve in $\AFBq$
  of Fig.~\ref{plot:vavap}. Here we see that it is possible to have
  large $\AFBq$ at low $q^2$, along with the suppression of $\DBRq$ in
  this region, as indicated by the Belle data.

\item It is possible to have $\DBRq$ consistent with the SM,
  simultaneously with a larger $\AFBq$ than the SM (up to 0.6) in the whole $q^2$ region.

\end{itemize}
The key point here is that, in order to reproduce the current
experimental data, one needs {\it both} $R_{V,A}$ and $R'_{V,A}$
couplings. They change $\AFBq$ appropriately in the low- and
high-$q^2$ regions, respectively. At present, the errors on the
measurements are quite large. However, if future experiments reproduce
the current central values with greater precision, this will put
important constraints on any NP model proposed to explain the data.

One NP model which contains VA operators (both $R_{V,A}$ and
$R'_{V,A}$) involves $Z'$-mediated FCNCs. A recent analysis
\cite{Z'FCNC} specifically notes that the measurement of $\AFBq$ can
be explained within this model. From the above analysis, we see that
this is one case of a more general result.

\subsection{Only SP new-physics operators}
\label{sec:sp}

\FIGURE[t]{
\epsfig{file=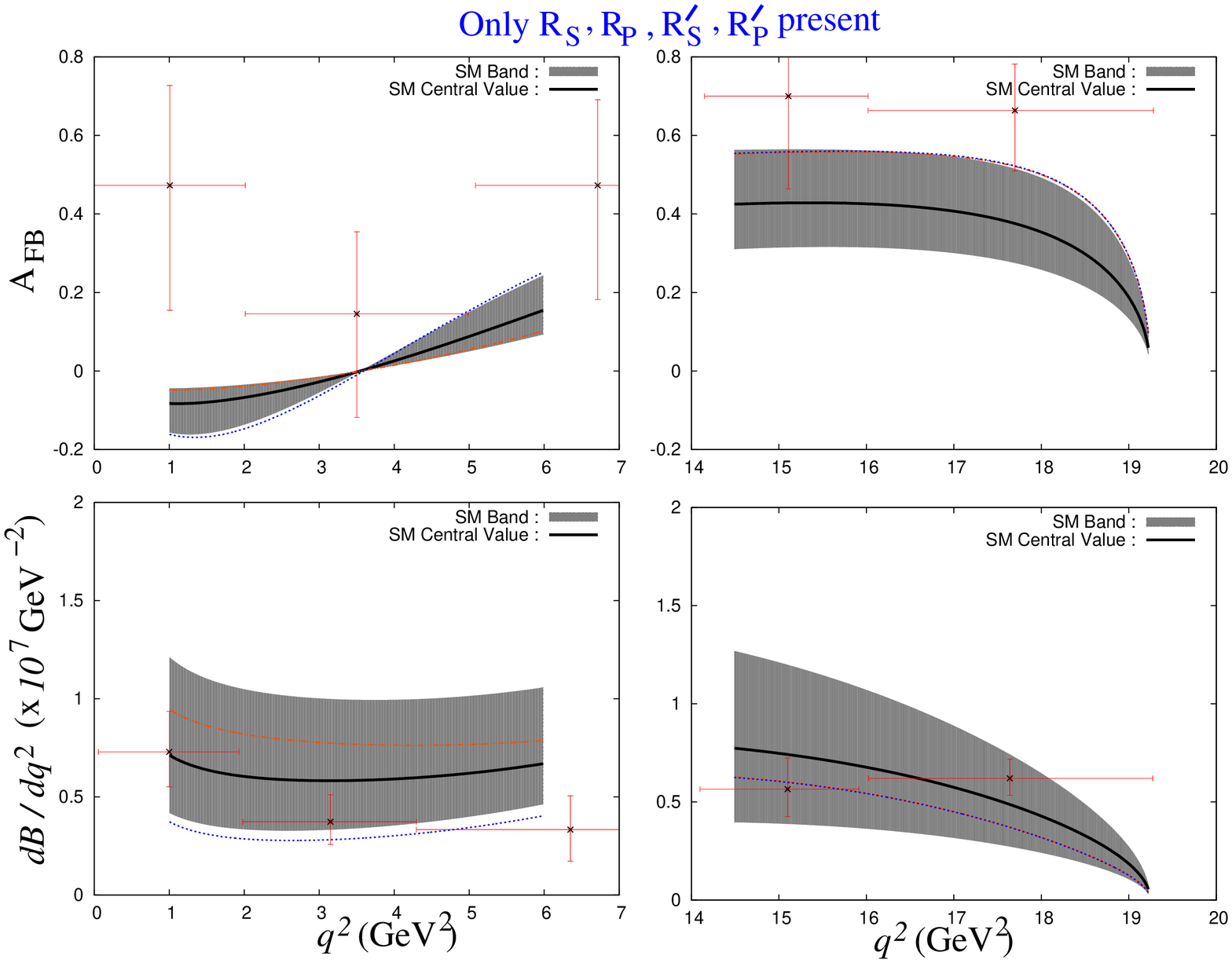,width=15cm}
\caption{The left (right) panels of the figure show $\AFBq$ and
  $\DBRq$ in the low-$q^2$ (high-$q^2$) region, in the scenario where
  both $R_{S,P}$ and $R'_{S,P}$ terms are present.}
    \label{plot:spspp}
}

{}From the discussion following Eq.~(\ref{Y-simplified}), NP involving
only SP operators is expected to decrease $\AFBq$.
Fig.~\ref{plot:spspp} shows the results when all the SP NP couplings,
$R_S, R_P, R'_S, R'_P$ are present. There we see that:
\begin{itemize}

\item The SP operators have unobservably small effects on $\AFBq$ 
and $\DBRq$.

\item There is always a SM-like zero crossing. 

\end{itemize}

Contribution to $\AFBq$ in this scenario can in principle come from
the terms $Y_{SP}$ and $Y_{SP-SM}$ in Eq.~(\ref{afb-expansion}).
However as as can be seen from Eq.~(\ref{Y-full-expansion}), $Y_{SP}$
vanishes identically while $Y_{SP-SM}$ is $\hat{m}_l$-suppressed.  In
addition, the couplings $R_{S,P}$ and $R'_{S,P}$ are strongly
constrained from the upper bound on $B({\bar B_s} \to
\mu^+\mu^-)$. For both of these reasons, these operators have a
negligible effect on $\AFBq$ and $\DBRq$.

\subsection{Only T new-physics operators}
\label{sec:t}

\FIGURE[t]{
\epsfig{file=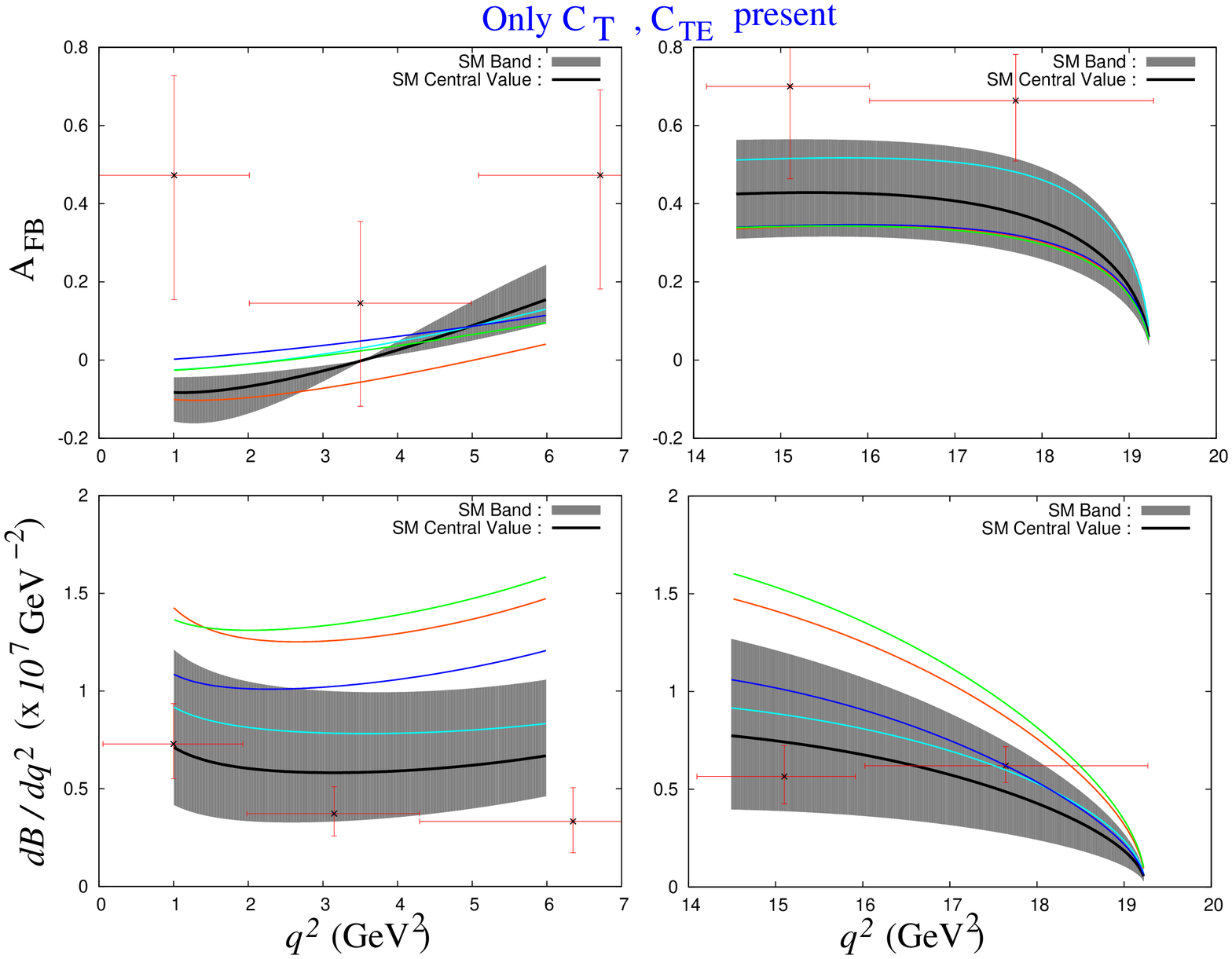,width=15cm}
\caption{The left (right) panels of the figure show $\AFBq$ and
    $\DBRq$ in the low-$q^2$ (high-$q^2$) region, when NP is present
    only in the form of tensor operators.}
    \label{plot:ot}
}

For the case where only tensor NP operators are added, $\AFBq$ is
expected to be suppressed, as the discussion following
Eq.~(\ref{Y-simplified}) suggests.  Fig.~\ref{plot:ot} shows the
results in this scenario.  The following remarks are in order:

\begin{itemize}

\item $\AFBq$ is in general suppressed in both the low- and
high-$q^2$ regions, as expected. 

\item The zero crossing can be anywhere in the entire $q^2$ range, or
  it may disappear altogether. Whenever it is present, it is always a
  positive crossing like the SM.  This shift of zero crossing shows
  that the $\hat{m}_l$-suppressed $Y_{SM{\hbox{-}}T}$ term in
  Eq.~(\ref{Y-full-expansion}) is important.  In the absence of this
  term, the zero crossing point would have remained the same as the
  SM.

\item $\DBRq$ is enhanced. The enhancement can be significant, up to a
  factor of 2.

\end{itemize}

Contributions to $\AFBq$ in this scenario are expected from the terms
$Y_{T}$ and $Y_{SM{\hbox{-}}T}$ in Eq.~(\ref{afb-expansion}).
However, as can be seen from Eq.~(\ref{Y-full-expansion}), $Y_{T}$
vanishes identically, while $Y_{SM{\hbox{-}}T}$ is
$\hat{m}_l$-suppressed.  On the other hand, the term $X_T$ has no such
suppression, and it contributes to $\Theta$, resulting in an
enhancement of $\DBRq$. The increased value of $\Theta$ also leads to
the suppression of $\AFBq$ in Eq.~(\ref{afb-expansion}).  In some
regions of parameter space, though, the contribution of the many terms
in $Y_{SM{\hbox{-}}T}$ is no longer negligible.  In such cases, the
zero crossing shifts and $\AFBq$ at low $q^2$ can become positive.

\subsection{Simultaneous SP and T new-physics operators }
\label{sec:sp-t}

\FIGURE[t]{
\epsfig{file=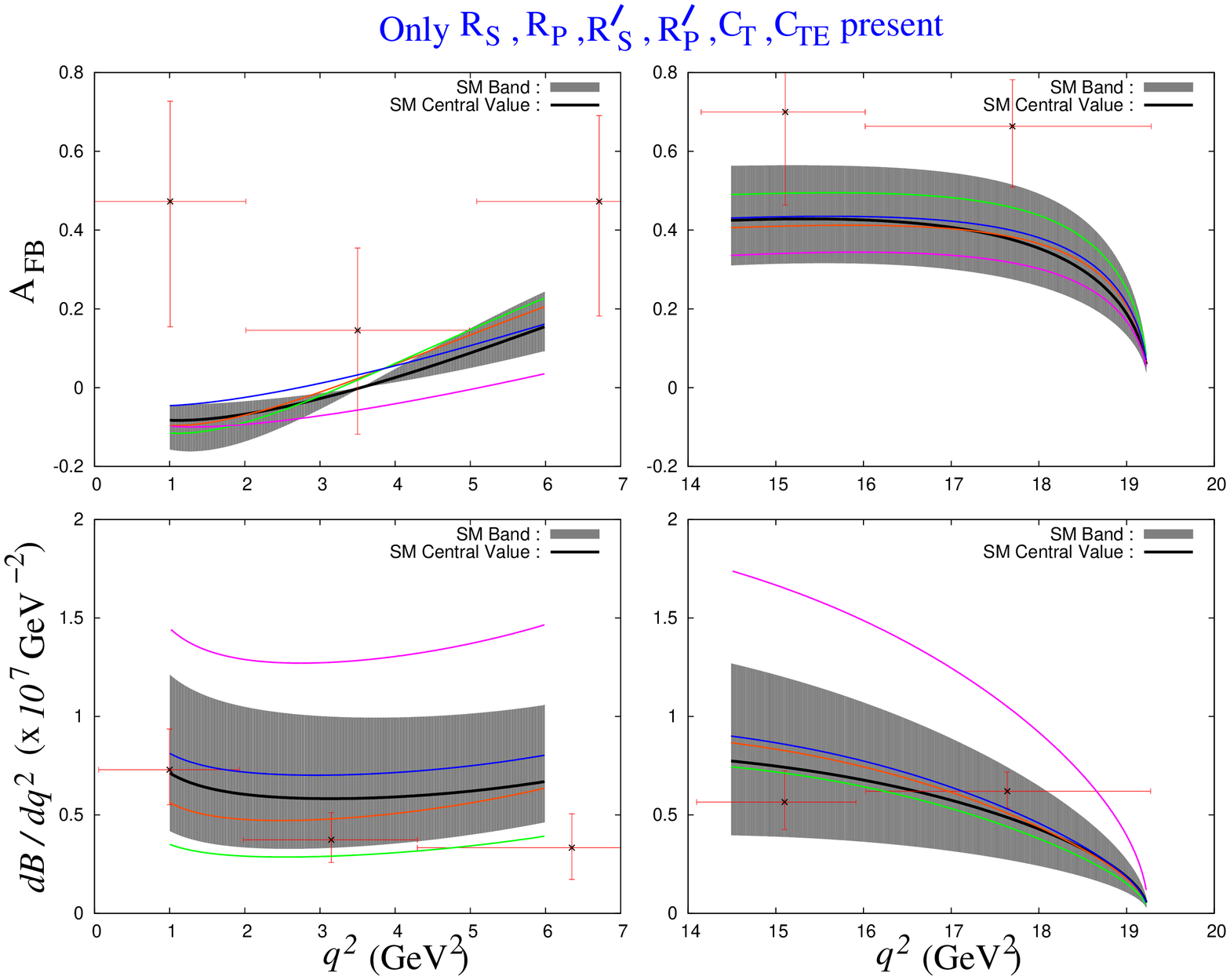,width=15cm}
\caption{The left (right) panels of the figure show $\AFBq$ and
    $\DBRq$ in the low-$q^2$ (high-$q^2$) region, when NP present is
    in the form of SP and T operators.}
    \label{plot:spsppt}
}

The discussion following Eq.~(\ref{Y-simplified}) suggests that if
both SP and T NP couplings are present simultaneously, there is the
possibility that $\AFBq$ is enhanced or changes sign.  In this
section, we quantitatively check if such an enhancement can take the
$\AFBq$ predictions closer to the current Belle measurements.  We take
the couplings $R_S, R_P, R'_S, R'_P, C_T, C_{TE}$ to be nonvanishing,
and show the results in Fig.~\ref{plot:spsppt}. From the figure, we
see the following:

\begin{itemize}

\item There is some parameter space of couplings where $\AFBq$ is
  positive everywhere, i.e. there is no zero crossing.

\item The absolute value of $\AFBq$ cannot be enhanced above the SM,
  except at very low $q^2$. Even here, the enhancement is very small.

\item $\DBRq$ is enhanced. The enhancement can be significant, up to a
  factor of 2.

\end{itemize}

Since the contribution to $\AFBq$ here comes from two terms, the
$\hat{m}_l$-suppressed (but not negligible) $Y_{SM{\hbox{-}}T}$ and
the not-$\hat{m}_l$-suppressed $Y_{SP{\hbox{-}}T}$ [see
  Eq.~(\ref{Y-full-expansion})], $\AFBq$ is now expected to be larger
than in the scenario with only T operators. Though this trend is
observed in general, the severe restrictions on the SP couplings do
not allow $\AFBq$ to become significantly more than the SM in
magnitude.  Still, $\AFBq$ can be influenced enough to cause a
vanishing of zero crossing and positive $\AFBq$ at low $q^2$.

Certain NP models can contribute to the SP-T mechanism, though there
are caveats.  In the MSSM, tensor operators in $b \to s\, \mu^+ \,
\mu^-$ are induced from photino and zino box diagrams. However, their
couplings are subleading in $\tan \beta$ with respect to the Higgs
penguins \cite{Bobeth:2007dw}. Tensor operators can also be induced in
leptoquark models by tree-level scalar leptoquark exchange (and a
Fierz transformation). However, the tensor couplings are suppressed by
the ratio of the Higgs vacuum expectation value and the scalar
leptoquark mass \cite{Davidson:1993qk}.

\subsection{Other combinations of VA, SP, and T operators}
\label{other-combinations}

The pattern of the effect of NP Lorentz structures on $\AFBq$ and
$\DBRq$ is now clear, so that the results from the remaining
combinations of operators can be discerned straightforwardly.  The
addition of VA operators allows for an enhanced $\AFBq$ and a moderate
enhancement of $\DBRq$.  The addition of SP operators does not affect
the results much due to the severe restrictions on the SP couplings.
The addition of T operators tends to enhance $\DBRq$ strongly and
decrease $\AFBq$ at the same time. In addition, we have found that
specific combinations of operators, such as SP-T, collaborate, with
the results approaching the experimental data.

\subsection{\boldmath Other new-physics sources that may affect $\AFBq$}
\label{O7-O7p}

In addition to the two NP mechanisms which have been found to
significantly affect $\AFBq$ (new VA operators, or an SP-T operator
combination), there are two additional mechanisms that can in
principle lead to the same effect.  We comment on them here.

In the first mechanism, NP is assumed to affect the ordinary dipole
operator $O_7 =\bar s\sigma^{\alpha\beta} P_R b \, F_{\alpha\beta}$.
In this case, the Wilson coefficient $C_7^{\rm eff}$ will be modified.
This will result in the shifting of the position of zero crossing, as
can be seen from Eq.~(\ref{zero}).  Now, there has been no hint of NP
in the radiative decays ${\bar B} \to X_s \gamma, {\bar K}^{(*)}
\gamma$, imposing strong constraints on $|C_7^{\rm eff}|$. Still, if
the effect of the NP is to simply reverse the sign of $C_7^{\rm eff}$,
then Eq.~(\ref{zero}) would not be fulfilled, and a positive $\AFBq$
would be produced for low $q^2$. However, this solution can be ruled
out at 3$\sigma$ from the decay rate of ${\bar B}\to X_s\ell^+\ell^-$
\cite{GHM}. This constraint can be evaded if the couplings $R_V$ and
$R_A$ are also taken to be nonzero. Thus, if there is NP in $O_7$
whose sole effect is to reverse the sign of $C_7^{\rm eff}$, {\it and}
the NP couplings $R_V$ and $R_A$ are present, it is possible to
reproduce the $\AFBq$ data. In other words, a great many things have
to happen perfectly for this mechanism to work. We consider this very
unlikely, and so consider this mechanism less plausible.

Another NP possibility, independent of those included in
Eq.~(\ref{NP:effHam}), is the addition of the chirally-flipped
operator $O_7^\prime= \bar s\sigma^{\alpha\beta} P_L b \,
F_{\alpha\beta}$.  The impact of this on $\AFBq$, together with other
observables, was studied in Ref.~\cite{Egede:2008uy}. There it was
found that $\AFBq$ does not significantly deviate from the SM
prediction if only this operator is introduced. We therefore exclude
the possibility of NP giving rise to $O_7^\prime$.

\section{Discussion and Summary
\label{summary}}

Motivated by the recent Belle measurement of the forward-backward
asymmetry $\AFBq$ in ${\bar B} \to {\bar K}^*\mu^+\mu^-$, indicating a
discrepancy with the SM, we calculate this quantity in the presence of
new physics (NP) in the low- and high-$q^2$ regions.  We perform a
systematic model-independent analysis, allowing for new vector-axial
vector (VA), scalar-pseudoscalar (SP) and tensor (T) couplings.  Using
the constraints on the new couplings from other related decays, we
determine how the NP affects $\AFBq$ and the differential branching
fraction $\DBRq$. This allows us to compare the effects of different
NP Lorentz structures.

The present Belle data \cite{BellePR} hint at a positive $\AFBq$ in
the whole $q^2$ region, i.e.\ no zero crossing as predicted by the SM.
Though we look for NP sources that can give rise to this feature, our
analysis is more general.  Indeed, the discrepancy with the SM
prediction is only at the $2\sigma$ level, and this could change with
more precise measurements in the future.  We therefore focus on
identifying unique features of all the sources of NP, and the patterns
of their effects on $\AFBq$ and $\DBRq$. We observe that the effects
on these two quantities are correlated, which could enable the
discrimination between different NP sources with future data.

We show, through analytical approximations and numerical calculations,
that two kinds of NP scenarios can give rise to a positive $\AFBq$ in
the whole $q^2$ range:
\begin{itemize}

\item
NP VA operators can enhance $\AFBq$ in the whole $q^2$ region and keep
its value positive. Both $R_{V,A}$ and $R'_{V,A}$ couplings are
necessary. The terms involving $R_{V,A}$ can make $\AFBq$ positive at
low $q^2$, while the terms involving $R'_{V,A}$ can increase $\AFBq$
above its SM value in the high-$q^2$ region. It is therefore possible
to very closely reproduce the Belle data. However, in general this
can also lead to a significant suppression of $\DBRq$. This is because
the VA operators can interfere with the SM operators without
$\hat{m}_l$ suppression, and a destructive interference in $\DBRq$
would tend to enhance $\AFBq$.  Still, $\AFBq$ values close to the
Belle data and $\DBRq$ consistent with the SM predictions are also
possible in this scenario.

\item
The T operators can influence $\AFBq$ in the low-$q^2$ region
sufficiently to change its sign and make it positive.  They still
cannot enhance the magnitude of the asymmetry significantly, since the
interference of these operators with the SM is $\hat{m_l}$-suppressed.
Moreover, the addition of these operators can only enhance $\DBRq$.
The simultaneous presence of SP operators allows interference terms
between SP and T operators that are not $\hat{m}_l$-suppressed, and
tends to take $\AFBq$ closer to the Belle data. However this
improvement is marginal, since the SP couplings are highly constrained
from the upper bound on $B({\bar B_s} \to \mu^+\mu^-)$.

\end{itemize} 

If the Belle anomaly remains in future measurements, the NP source has
to be one of the above two (or the less plausible mechanism involving
a conspiracy between $O_7$, $R_V$ and $R_A$ operators to flip the sign
of $O_7$). We will be able to distinguish between them through the
following observations:
\begin{itemize}

\item $\AFBq$ at high $q^2$: the scenarios with only T operators or an
  SP-T combination cannot give rise to an enhanced $\AFBq$ at high
  $q^2$. Such a situation necessarily requires VA operators, in
  particular involving the couplings $R'_{V,A}$.

\item Correlation with $\DBRq$: the T-only or SP-T scenarios cannot in general
  suppress $\DBRq$ much below its SM value, while the VA scenario will
  be able to do this.

\item Correlation with $\AFBq$ in ${\bar B} \to {\bar K} \mu^+\mu^-$:
  within the SM, $\AFBq$ in ${\bar B} \to {\bar K} \mu^+ \mu^-$ is
  consistent with zero since the hadronic matrix element for the $B \to K$
  transition does not get any contribution from axial-vector current. For the same
  reason, new VA operators cannot contribute to $\AFBq$ in ${\bar B}
  \to {\bar K}\mu^+\mu^-$.  Hence, the only possible contribution to
  the asymmetry can be from SP or T operators. In
  Ref.~\cite{ADSKmumu}, it was shown that, if the NP is only in the
  form of SP or T operators, then the additional contribution to
  $\AFBq$ is proportional to the lepton mass and hence is highly
  suppressed. However, if both SP and T operators are present
  simultaneously, then $\AFBq$ in ${\bar B} \to {\bar K}\mu^+\mu^-$
  can be as large as 15$\%$.  Thus, if the SP-T scenario is
  responsible for $\AFBq$ in ${\bar B} \to {\bar K}^* \mu^+\mu^-$, we
  will also see a large $\AFBq$ in ${\bar B} \to {\bar K}
  \mu^+\mu^-$. This measurement may also help distinguishing between
  the T-only and SP-T scenarios.

\end{itemize}

Even if the Belle anomaly does not persist, or shows some other
features, our analysis enables us to recognize patterns of the
effects of VA, SP and T operators on $\AFBq$ and $\DBRq$:
\begin{itemize}

\item New VA operators can interfere with the SM operators without
  $\hat{m}_l$ suppression, and hence are the only ones that can reduce
  $\DBRq$ substantially below the SM prediction.  They can also
  interfere among themselves and with the SM operators to give rise to
  a large magnitude for $\AFBq$, with either sign. They can influence
  the zero crossing point $q_0$ by changing the SM relation between
  $C_9^{\rm eff}$ and $C_7^{\rm eff}$ that determines its value at LO.

\item SP operators always tend to enhance $\DBRq$, since their
  interference terms with the SM or VA operators are
  $\hat{m}_l$-suppressed.  The contribution of only SP operators to
  $\AFBq$ is also $\hat{m}_l$-suppressed. Moreover, the magnitudes of
  the SP couplings are severely constrained from the upper bound on
  $B({\bar B_s} \to \mu^+\mu^-)$. Therefore, the addition of only SP
  operators does not significantly affect either $\DBRq$ or $\AFBq$.

\item T operators also always tend to enhance $\DBRq$, owing to the
  $\hat{m}_l$ suppression of their interference with the SM operators.
  The enhancement can be rather strong -- up to a factor of 2 -- since
  the constraints on the couplings are relatively weak.  If the NP is
  only in the form of T operators, then the contribution to $\AFBq$ is
  also $\hat{m}_l$-suppressed, though the presence of many such terms
  mean that the total contribution is not insignificant.  The net
  effect is that the scenario with only T operators tends to show a
  large $\DBRq$ enhancement and $\AFBq$ suppression.

\item The simultaneous presence of SP and T operators gives rise to a
  qualitatively new feature. Though the $\DBRq$ can be enhanced, one also
  gets an interference term between the SP and T operators that is not
  $\hat{m}_l$-suppressed. As a result, a substantial effect on $\AFBq$
  is possible, though a large effect is not possible due to the
  restriction on the SP couplings.

\end{itemize}

\bigskip
\noindent
{\bf Acknowledgments}:
We thank Marco Musy, F. Mescia and Mitesh Patel for useful discussions. This work
was financially supported by NSERC of Canada (AKA, MN, AS, DL).  JM
acknowledges financial support from the Research Projects
CICYT-FEDER-FPA2008-01430, SGR2005-00916, PORT2008.

\begin{appendix}

\section{\boldmath Analytical Calculation of $\AFBq$ and $\DBRq$
\label{appendix-1}}

The decay amplitude for $\bkll$ [Eq.~(\ref{bkllampNP})] is
written in terms of matrix elements of the quark
operators. These are \cite{TheobsllSM}
\bea
\langle {\bar K}^*(p_{K^*},\epsilon)|\bar{s}\gamma_\mu (1 \pm
\gamma_5)b|{\bar B}(p_B) \rangle & ~=~ & \mp~iq_{\mu}
\frac{2m_{K^*}}{q^2} \, \epsilon^{*} \cdot q \, \bigg[
A_3(q^2)-A_0(q^2) \bigg] \phantom{patchthisup} \nn \\
&& \hskip-4.5truecm \pm~i\epsilon_{\mu}^{*}(m_B+m_{K^*})
A_1(q^2) \mp~i(p_B+p_{K^*})_{\mu} \, \epsilon^{*} \cdot q
\, \frac{A_2(q^2)}{(m_B+m_{K^*})} \nn \\
&& \hskip-4.5truecm
-~\epsilon_{\mu\nu\lambda\sigma} \epsilon^{*\nu}
p^{\lambda}_{K^*} q^{\sigma} \frac{2V(q^2)}{(m_B+m_{K^*})} ~, 
\label{me1}
\eea
where
\beq
A_3(q^2)~=~\frac{m_B+m_{\kstar}}{2m_{\kstar}}A_1(q^2)-\frac{m_B-m_{\kstar}}{2m_{\kstar}}A_2(q^2)\; ,
\label{a3-a1-a2}
\eeq
\bea
\langle {\bar K}^*(p_{K^*},\epsilon)|\bar{s}\sigma_{\mu\nu}b|{\bar B}(p_B)
\rangle & ~=~ & i\epsilon_{\mu\nu\lambda\sigma} \bigg\{ -
T_1(q^2)\epsilon^{*\lambda} (p_B+p_{K^*})^{\sigma}  \nn \\
&& \hskip-2truecm +~\frac{ (m_B^2-m_{K^*}^2) }{q^2}
\bigg(T_1(q^2)-T_2(q^2) \bigg)\epsilon^{*\lambda} q^{\sigma}
\phantom{patchupthisequation}
 \nn \\
&& \hskip-4truecm -~\frac{2}{q^2} \bigg(T_1(q^2)-T_2(q^2)
-\frac{q^2}{(m_B^2-m_{K^*}^2)}~T_3(q^2) \bigg) \epsilon^{*}
\cdot q \, p^{\lambda}_{K^*} q^{\sigma} \bigg\}\; ,
\label{me2}
\eea
\bea
\langle {\bar K}^*(p_{K^*},\epsilon)|\bar{s}i\sigma_{\mu\nu}q^\nu (1
\pm \gamma_5)b|{\bar B}(p_B) \rangle & ~=~ & 2
\epsilon_{\mu\nu\lambda\sigma} \epsilon^{*\nu}
p^\lambda_{K^*} q^\sigma ~ T_1 (q^2) \nn \\
&& \hskip-2truecm \pm~i \bigg\{
\epsilon_{*\mu}(m_B^2-m_{K^*}^2)-(p_B+p_{K^*})_\mu
\, \epsilon^{*} \cdot q
\,  \bigg\}~T_2(q^2) \nn \\
&& \hskip-2truecm \pm~i \, \epsilon^{*} \cdot q
\, \bigg\{ q_\mu -
\frac{(p_B + p_{K^*})_\mu q^2}{(m_B^2-m_{K^*}^2)} \bigg\} ~
T_3(q^2)\; ,
\label{me3} 
\eea
\bea
\langle {\bar K}^*(p_{K^*},\epsilon)|\bar{s}(1\pm\gamma_5)b|{\bar B}(p_B)
\rangle & ~=~ & \mp~2i \frac{m_{K^*}}{m_b} \,
\epsilon^{*} \cdot q \,A_0(q^2) \;. \phantom{patchupthisequational}
\label{me4} 
\eea 
Here we have neglected the strange-quark mass. The matrix
elements are functions of 7 unknown form factors:
$A_{0,1,2}(q^2)$, $V(q^2)$, $T_{1,2,3}(q^2)$.

Using the above matrix elements, the decay amplitude for
$\bkll$ can be written as
\bea
{\cal M} & ~=~ & \frac{\alpha G_F}{4\sqrt{2}\pi}
V^*_{ts} V_{tb} \Bigg[ (\bar{u}(p_{-})\gamma^{\mu} v(p_{+}))
\times \\
&& \hskip-0.4truecm \bigg\{ -2 A
\epsilon_{\mu\nu\lambda\sigma} \epsilon^{*\nu}p_{K^*}^\lambda
q^\sigma - i B \epsilon_\mu^* + i C \, \epsilon^{*} \cdot q
\, (p_B + p_{K^*})_\mu + i D \, \epsilon^{*} \cdot q \, q_\mu
\bigg\} \nn \\
&& \hskip-0.4truecm +~(\bar{u}(p_{-}) \gamma^{\mu}\gamma_5
v(p_{+})) \times \nn\\
&& \hskip2truecm
\bigg\{ -2 F_1 \epsilon_{\mu\nu\lambda\sigma}
\epsilon^{*\nu} p_{K^*}^{\lambda}q^{\sigma} - i
F\epsilon_{\mu}^{*} + iG \, \epsilon^{*} \cdot q \,
(p_B+p_{K^*})_{\mu} + i H \, \epsilon^{*} \cdot q \, q_{\mu}
\bigg\} \phantom{spa} \nn \\
&& \hskip-0.4truecm +~ i B_1 (\bar{u}(p_{-}) v(p_{+}) ) \,
\epsilon^{*} \cdot q \, + i B_2 ~(\bar{u}(p_{-}) \gamma_5
v(p_{+})) \, \epsilon^{*} \cdot q  \nn \\
&& \hskip-0.4truecm +~8
C_{TE}~(\bar{u}(p_{-})\sigma_{\mu\nu} v(p_{+})) ~\bigg\{ -2
T_1\epsilon^{*\mu}(p_B+p_{K^*})^{\nu} +
B_3\epsilon^{*\mu}q^{\nu} - B_4 \, \epsilon^{*} \cdot q \,
p_{K^*}^{\mu}q^{\nu} \bigg\} \nn \\
&& \hskip-0.4truecm +~2 i
C_{T}~\epsilon_{\mu\nu\lambda\sigma}
~(\bar{u}(p_{-})\sigma^{\mu\nu} v(p_{+})) ~\left\{ - 2
T_1\epsilon^{*\lambda}(p_B+p_{K^*})^{\sigma} +
B_3\epsilon^{*\lambda}q^{\sigma} - B_4 \, \epsilon^{*} \cdot
q \, p_{K^*}^{\lambda}q^{\sigma} \right\} \Bigg], \nn
\eea
with the quantities $A, B, C, D, F_1, F, G, H$, that are relevant for
VA interactions, defined as
\bea
A & ~=~ & 2(C_{9}^{eff}+R_V+R_{V}^\prime
)\, \frac{V(q^2)}{m_B(1+\hat{k})} + \frac{4{m_b}C_{7}^{eff} T_1(q^2)}{q^2}\,~, \nn \\
B & ~=~ & 2(C_{9}^{eff}+R_V-R_{V}^\prime) \, m
_B(1+\hat{k})A_1(q^2)
+4m_bC_{7}^{eff}(1-\h{k}^2) \, \frac{T_2(q^2)}{(q^2/m_B^2)}\, ~, \nn \\ 
C & ~=~ & 2(C_{9}^{eff}+R_V-R_{V}^\prime
) \, \frac{A_2(q^2)}{m_B(1+\hat{k})}
+\frac{4{m_b}C_{7}^{eff}}{q^2}\bigg[T_2(q^2)+\frac{(q^2/m_B^2)}{(1-\hat{k}^2)}
  \, T_3(q^2)\bigg] \,
~, \nn \\ 
D & ~=~ &
\frac{4\hat{k}}{m_B}(C_{9}^{eff}+R_V-R_{V}^\prime) \, \frac{A_3(q^2)-A_0(q^2)}{(q^2/m_B^2)} -
\frac{4{m_b}C_{7}^{eff}T_3(q^2)}{q^2} \, ~, 
\nn \\ 
F_1 & ~=~ &
(C_{10}+R_A+R_{A}^\prime) \, \frac{2V(q^2)}{m_B(1+\hat{k})} \,
~, \nn \\ 
F & ~=~ & 2(C_{10}+R_A-R_{A}^\prime) \, m_B(1+\hat{k})A_1(q^2) \,
~, 
\nn \\
G & ~=~ &
(C_{10}+R_A-R_{A}^\prime) \, \frac{2 A_2(q^2)}{m_B(1+\hat{k})} \,
~, \nn \\ 
H & ~=~ &
\frac{4\hat{k}}{m_B}(C_{10}+R_A-R_{A}^\prime) \, \frac{A_3(q^2)-A_0(q^2)}{(q^2/m_B^2)} \; .
\eea
The quantities $B_{1,2,3,4}$, relevant for SP and T interactions,
are defined as
\bea
B_1 & ~=~ &  -4(R_S - R_{S}^\prime) \,
\frac{\hat{k}}{(m_b/m_B)} \, A_0(q^2)\,
~, \nn \\ 
B_2 & ~=~ & -4(R_P - R_{P}^\prime) \,
\frac{\hat{k}}{(m_b/m_B)} \, A_0(q^2) \,
~, \nn \\ 
B_3 & ~=~ & 2(1-\hat{k}^2) \, \frac{T_1(q^2)-T_2(q^2)}{(q^2/m_B^2)} \, ~, \nn \\
B_4 & ~=~ &
\frac{4}{q^2}\bigg(T_1(q^2)-T_2(q^2)-\frac{(q^2/m_B^2)}{(1-\hat{k}^2)}
\, T_3(q^2)\bigg)\, ,
\eea
where $q=(p_{+}+p_{-})$ and $\hat{k}\equiv m_{K^*}/m_B$.

The double differential decay rate is given by
\beq
\frac{d^2\Gamma}{dq^2 d\cos\theta} = \frac{1}{2E_B}
\frac{2 v \sqrt{\lambda}}{(8 \pi)^3} |M|^2 \; ,
\eeq
where $v \equiv \sqrt{1 - 4 m_l^2/q^2}$. 
Here $\lambda \equiv 1 + \hat{r}^2 + z^2 - 2 (\hat{r} +z)-2\hat{r} z$,
with $\hat{r} \equiv m_{K^*}^2/m_B^2$ and $z \equiv q^2/m_B^2$.
This leads to the differential branching ratio:
\begin{equation}
\frac{dB}{dq^2} = \frac{G^2 \alpha^2}{2^{14}} \frac{1}{\pi^5}
|V_{tb}V_{ts}^{*}|^2 m_{B} \tau_B \sqrt{\lambda} \, \Theta ~,
\end{equation}
where $\tau_B$ is the lifetime of $B$ meson.
The quantity $\Theta$ has the form
\beq
\Theta = \frac{1}{3\h{r}}\Big[X_{SP}+X_{VA}+X_{T}+X_{SP{\hbox{-}}VA}+X_{SP{\hbox{-}}T}
+X_{VA{\hbox{-}}T}\Big] \; ,
\eeq
where the complete expressions for the $X$ terms are:
\begin{eqnarray}
X_{SP}&=&3 \left|B_{1}\right|^2 m_{B}^2 z v^2 \lambda + 3 \left|B_{2}\right|^2 m_{B}^2 z \lambda \nonumber \\
X_{VA}&=&-8 \left|A\right|^2 m_{B}^4 \h{r} z \left(v^2-3\right) \lambda -\left|B\right|^2 \left(v^2-3\right)(12 \h{r} z+\lambda ) 
-\left|C\right|^2 m_{B}^4 \left(v^2-3\right) \lambda ^2  \nn\\
&&+~\left|F\right|^2 \Bigg(24 \h{r} z v^2-\left(v^2-3\right) \lambda \Bigg)+16 \left|F_1\right|^2 m_{B}^4 \h{r} z v^2 \lambda  \nonumber \\
&& +~\left|G\right|^2 m_{B}^4 \lambda 
\Bigg(-6 (\h{r}+1) z \left(v^2-1\right)+3 z^2 \left(v^2-1\right)-\left(v^2-3\right) \lambda \Bigg) \nonumber \\
&& -~3 \left|H\right|^2 m_{B}^4 z^2 \left(v^2-1\right) \lambda + 2{\rm Re}\left(F G^*\right) m_{B}^2 \lambda   
\Bigg(-\h{r} \left(v^2-3\right)+(2 z+1) v^2-3\Bigg) \nonumber \\
&& +~6 {\rm Re}\left(F H^*\right) m_{B}^2 z \left(v^2-1\right)\lambda + 6 {\rm Re}\left(G H^*\right) m_{B}^4 (\h{r}-1) z \left(v^2-1\right) \lambda
\nonumber \\
&& -~2{\rm Re}\left(B C^{*}\right) m_{B}^2 \left(v^2-3\right) \lambda 
(\h{r}+z-1)  \nonumber
\eea
\bea
X_{T}&=&16 m_{B}^2 \left|C_{TE}\right|^2 \Bigg\{-4 B_{3}^2 z \left(2 v^2-3\right) (12 \h{r} z+\lambda )
\nn\\
&& -~4 B_{3} z \left(2 v^2-3\right)
 \lambda  \left(B_{4} m_{B}^2 (\h{r}+z-1)+44 T_{1}\right) \nonumber \\
&& -~192 B_{3} z T_{1} \left(2 v^2-3\right) \left(2 \h{r} z+\h{r}-(z-1)^2\right)-B_{4}^2 m_{B}^4 z \left(2 v^2-3\right) \lambda ^2
 \nonumber \\
&& -~8 T_{1} \lambda\left.\Bigg(B_{4} m_{B}^2 z \left(2 v^2-3\right) (3 \h{r}-z+1)+2 T_{1} \left(8 \h{r} \left(v^2-3\right)+25 z \left(2 v^2-3\right)
\right)\Bigg) \right.\nonumber \\
&& -192 z T_{1}^2 \left(2 v^2-3\right) \left(3 \h{r} (z+2)-2 (z-1)^2\right)\Bigg\} \nonumber \\
&& +~4 m_{B}^2 \left|C_T\right|^2 \Bigg\{4 \lambda  
\Bigg(v^2 \Bigg(2 z T_{1} \left(22 B_{3}+B_{4} m_{B}^2 (3 \h{r}-z+1)\right) \nn\\
&& +~B_{3} z \left(B_{3}+B_{4} m_{B}^2 (\h{r}+z-1)\right) 
+4 T_{1}^2 (25 z-8 \h{r})\Bigg)+96 \h{r} T_{1}^2\Bigg) \nonumber \\
&& +~48 z v^2 \Bigg((B_{3}+2 T_{1}) (B_{3} \h{r} z +
2 T_{1} (z (3 \h{r}-2 z+4)-2))+
4 \h{r} T_{1} (B_{3}+6 T_{1})\Bigg) \nonumber \\
&& +~B_{4}^2 m_{B}^4 z v^2 \lambda ^2\Bigg\} \nonumber\\
X_{SP{\hbox{-}}VA}&=& 12 m_{B} \h{m_{l}} \lambda  {\rm Re}\left(B_{2}^* F\right)-12 m_{B}^3 \h{m_{l}} (\h{r}-1) \lambda
{\rm Re}\left(B_{2}^* G\right)+12 m_{B}^3 \h{m_{l}} z \lambda  {\rm Re}\left(B_{2}^* H\right) \nonumber \\
X_{SP{\hbox{-}}T}&=& 0 \nonumber \\
X_{VA{\hbox{-}}T}&=& 768 m_{B}^3 \h{m_{l}} \h{r} T_{1} \lambda  {\rm Re}\left(A^* C_T\right)+48 m_{B} \h{m_{l}}
{\rm Re}\left(B^* C_{\rm{TE}}\right) \times \nn\\
&& \hskip-1truecm
\Bigg(2 B_{3} (12 \h{r} z+\lambda )+B_{4} m_{B}^2 \lambda  (\h{r}+z-1)+ 
 T_{1} \left(96 \h{r} z+48 \h{r}-48 z^2+96 z+44 \lambda -48\right)\Bigg) \nonumber \\
&& +48 m_{B}^3 \h{m_{l}} \lambda  {\rm Re}\left(C^* C_{TE}\right) \Bigg(2 B_{3} (\h{r}+z-1)+B_{4} m_{B}^2 \lambda +
4 T_{1} (3 \h{r}-z+1)\Bigg)
\label{X-full-expansion}
\end{eqnarray}

Note that here, we do not differentiate between the contributions
from the SM and the new VA operators.
The forward-backward asymmetry can also be written in the form
\beq
\AFBq =2 m_{B} \frac{\sqrt{\lambda}}{\hat{r} \Theta}
\Big[Y_{SP}+Y_{VA}+Y_{T}+Y_{SP{\hbox{-}}VA}+Y_{SP{\hbox{-}}T}+Y_{VA{\hbox{-}}T}\Big] \;,
\eeq
with the complete expressions for the $Y$ terms given as
\begin{eqnarray}
Y_{SP}&=& 0 \; ,\nonumber \\ 
Y_{VA}&=& -4 m_{B} \h{r} z {\rm Re}\Big(A^* F+B^* F_1\Big) \; ,\nonumber \\
Y_{T}&=& 0 \; ,\nonumber 
\end{eqnarray}
\begin{eqnarray}
Y_{SP{\hbox{-}}VA}&=& \h{m_l} (\h{r}+z-1) {\rm Re}\left(B^* B_{1}\right)+m_{B}^2 \h{m_l} \lambda  {\rm Re}\left(B_{1}^* C\right) \; , \nonumber \\
Y_{SP{\hbox{-}}T}&=& m_B z {\rm Re}(2B_{1}^* C_{\textit{TE}}+ B_{2}^*C_T)
\Big(2 B_{3} (\h{r}+z-1)+B_{4} m_{B}^2 \lambda +4 T_{1} (3 \h{r}-z+1)\Big) \; ,
\nonumber \\
Y_{VA{\hbox{-}}T}&=& 2 {\rm Re}\left(F^* C_T\right) \h{m_l}  \Big(2 B_{3} (\h{r}+z-1)+B_{4} m_{B}^2 \lambda +T_{1} (44 \h{r}-4 z+4)\Big)  \nonumber \\
&& -2 {\rm Re}\left(G^* C_T\right) m_{B}^2 \h{m_l}  \Big(2 B_{3} \left(3 \h{r} z-z^2+z+\lambda \right)+B_{4} m_{B}^2 (\h{r}-1) \lambda \nonumber \\
&&  +4 T_{1} \left(5 \h{r} z+4 \h{r}-3 z^2+7 z+3 \lambda -4\right)\Big) \nonumber \\
&& + 2 {\rm Re}\left( H^* C_T\right) m_{B}^2 \h{m_l} z  \Big(2 B_{3} (\h{r}+z-1)+B_{4} m_{B}^2 \lambda +4 T_{1} (3 \h{r}-z+1)\Big) \nonumber \\
&& -64 {\rm Re}\left(F_1^* C_{\textit{TE}}\right) m_{B}^2 \h{m_l}  \Big(B_{3} \h{r} z+2 T_{1} \left(2 \h{r} z+\h{r}-(z-1)^2+\lambda \right)\Big) \; .
\label{Y-full-expansion}
\end{eqnarray}

\end{appendix}

\end{document}